\documentclass[article]{jpconf}

\usepackage[english]{babel}
\usepackage[utf8]{inputenc}
\usepackage[T1]{fontenc}

\usepackage[a4paper]{geometry}

\usepackage{amsmath}
\usepackage[colorlinks=true, allcolors=blue]{hyperref}
\usepackage{bbold}
\usepackage[shortlabels]{enumitem}
\usepackage{braket}
\usepackage{amsmath,amssymb}
\usepackage{pifont}
\def\club{\ding{168}}
\def\diamond{\color{red}\ding{169}}
\def\heart{\color{red}\ding{170}}
\def\spade{\ding{171}}

\usepackage{skak}

\usepackage{amsfonts,amsthm}
\usepackage{bm,bbm}
\usepackage{geometry}
\usepackage{mathtools}
\usepackage{color}
\usepackage{setspace}
\usepackage[dvipsnames]{xcolor}
\usepackage{hyperref}
\usepackage{tensor}
\usepackage{hyperref}
\usepackage{courier}
\usepackage{skak}
\usepackage{xcolor,colortbl}
\usepackage{cite}

\usepackage{stackengine,xcolor}

\newcommand\shadowfy[1]{\expandafter\shadowfypars#1\par\relax\relax}
\long\def\shadowfypars#1\par#2\relax{%
  \ifx#1\relax\else
    \shadowfywords#1 \relax\relax%
  \fi%
  \ifx\relax#2\else\par\shadowfypars#2\relax\fi%
}
\def\shadowfywords#1 #2\relax{%
  \ifx#1\relax\else
    \shadowfyletters#1\relax\relax%
  \fi%
  \ifx\relax#2\else\ \shadowfywords#2\relax\fi%
}
\def\shadowfyletters#1#2\relax{%
  \shadow{#1}%
  \ifx\relax#2\else\shadowfyletters#2\relax\fi}

\newlength\shadowHoffset
\newlength\shadowVoffset
\def\primarycolor{white}
\def\secondarycolor{black}

\def\shadow#1{\kern\shadowHoffset%
  \raisebox{\shadowVoffset}{%
  \stackinset{c}{-\shadowHoffset}{c}{-\shadowVoffset}%
  {\textcolor{\primarycolor}{#1}}{\textcolor{\secondarycolor}{#1}}}%
  \kern-\shadowHoffset}

\definecolor{orange}{rgb}{1,0.7,0}
\definecolor{myred}{rgb}{0.6,0.2,0.2}

\usepackage{chessfss}
\usepackage{adjustbox}
\setfigfontfamily{berlin}

\newcommand{\pawnB}[1][1.3ex]{%
\adjustbox{Trim=4.3pt 2.6pt 4.3pt 0pt,width=#1,margin=0.2ex 0ex 0.2ex 0ex}{\BlackPawnOnWhite}%
}%
\newcommand{\rookB}[1][1.58ex]{%
\adjustbox{Trim=3.2pt 2.2pt 3.2pt 0pt,width=#1,raise=0ex,margin=0.1ex 0ex 0.1ex 0ex}{\BlackRookOnWhite}%
}%
\newcommand{\knightB}[1][1.85ex]{%
\adjustbox{Trim=2.3pt 2.35pt 2.5pt 0pt,width=#1,raise=-0.03ex,margin=0.14ex 0ex 0.14ex 0ex}{\BlackKnightOnWhite}%
}%
\newcommand{\bishopB}[1][1.79ex]{%
\adjustbox{Trim=2.3pt 2pt 2.3pt 0pt,width=#1,raise=-0.12ex,margin=0.1ex 0ex 0.1ex 0ex}{\BlackBishopOnWhite}%
}%
\newcommand{\queenB}[1][2.05ex]{%
\adjustbox{Trim=1.2pt 2.2pt 1.2pt 0pt,width=#1,raise=-0.08ex,margin=0.1ex 0ex 0.1ex 0ex}{\BlackQueenOnWhite}%
}%
\newcommand{\kingB}[1][1.95ex]{%
\adjustbox{Trim=2pt 2pt 2pt 0pt,width=#1,raise=-0.06ex,margin=0.13ex 0ex 0.13ex 0ex}{\BlackKingOnWhite}%
}%

\begin{document}

\title{9 $\times$ 4 = 6 $\times$ 6: \  Understanding the  quantum solution 
       to Euler's problem of 36 officers}

\author{Karol \.Zyczkowski$^{1,2}$, Wojciech Bruzda$^1$, Grzegorz Rajchel-Mieldzio\'c$^{3,2}$, Adam Burchardt$^{4,1}$, Suhail Ahmad
Rather$^5$, Arul Lakshminarayan$^5$}
\address{$^1$ Institute of Theoretical Physics, Jagiellonian University, ul. {\L}ojasiewicza 11, 30-348 Krak\'ow, Poland}
\address{$^2$ Center for Theoretical Physics, Polish Academy of Sciences, 02-668 Warszawa, Poland}
\address{$^3$ ICFO-Institut de Ciencies Fotoniques, The Barcelona Institute of Science and Technology, Av. Carl Friedrich Gauss 3, 08860 Castelldefels (Barcelona), Spain}
\address{$^4$ QuSoft, CWI and University of Amsterdam, Science Park 123, 1098 XG Amsterdam, the Netherlands}
\address{$^5$ Department of Physics, Indian Institute of Technology Madras, Chennai 600036, India}

\date{April 29, 2022}

\begin{abstract}
 The famous combinatorial problem of Euler concerns an arrangement of $36$ officers
from six different regiments in a $6 \times 6$ square array. Each regiment consists of six officers each belonging to one of six ranks. The problem, originating from 
Saint Petersburg,
requires that each row and each column of the array contains only one officer of a given rank and given regiment.
Euler observed that such a configuration does not exist.
In recent work, we constructed a solution to a quantum version of this problem assuming that the officers correspond to quantum states and can be entangled.
In this paper, we explain the solution which 
is based on a partition of 36 officers into nine groups, each with four elements.
The corresponding quantum states are locally equivalent to maximally
entangled two-qubit states, hence each officer is 
entangled with at most three out of his 35 colleagues. 
The entire quantum combinatorial design involves 9 Bell bases
in nine complementary 4--dimensional subspaces. \ \ 
\end{abstract}


\section{Introduction}

Designs composed of elements of a finite set arranged with certain symmetry and balance have been known since antiquity. Nowadays, they are commonly referred to as \textit{combinatorial designs} \cite{CD07}, 
mostly developed in the $18^{\rm th}$ century along with the general growth of combinatorics. Some of them, such as magic squares and Sudoku grids, offer mathematical recreations. However, the $20^{\rm th}$ century brought to light several applications of combinatorics to real-world problems. The theory of combinatorial designs is applicable to the area of design of biological experiments, tournament scheduling, algorithm design and analysis, group testing, and cryptography \cite{CD07}. 

The notion of combinatorial designs can be quantized. 
In a 1999 Ph.D. thesis of Zauner \cite{Za99}, it was proposed to
analyze particular configurations of quantum states 
belonging to a finite-dimensional Hilbert space, $|\psi_j\rangle \in {\cal H}_N$,
called {\sl quantum designs}, which satisfy certain symmetry conditions.
Observe that constructing a classical design
involves working with a finite set of objects and their transformations under the discrete permutation group. 
In a quantum design, the set of states of size $N$ is continuous and is governed by the unitary group $\mathbb{U}(N)$. 

Quantum combinatorial designs allow for various interpretations
and offer several applications of quantum theory.
For instance, {\sl mutually unbiased bases} (MUBs) 
\cite{Iv81,Wo89,KR04,DEBZ10,MCST21}
and {\sl symmetric informationally complete positive operator valued measures} (SIC POVMs) \cite{Za99,RBSC04,FHS17,GS17,AB19}
can provide schemes of quantum measurements with 
particularly appealing properties.
A constellation of pure states of a simple system 
forms a {\sl quantum $t$--design} \cite{AE07}
if the average of any polynomial function of order $t$
over the constellation agrees with the average over the entire space
of pure quantum states with respect to the natural unitarily invariant measure.
In particular, MUBs and SICs form a special case of quantum $2$--designs,
while designs of a higher degree $t$ are also known \cite{KG15}.

In the case of multipartite systems,
examples of often analyzed structures include symmetric states with high geometric entanglement
\cite{MGP10} and designs in composite spaces constructed out
of iso-entangled states \cite{ZYE10,CGGZ20}.
Another related problem is the construction of
{\sl absolutely maximally entangled} (AME) states
of a system composed of $N=2k$ parties, 
such that the state is maximally entangled for any symmetric splitting into 
two parties of $k$ subsystems each \cite{Sc04,BP+07,FFPP08,HCLRL12}. AME states are resources for various quantum information protocols like parallel quantum teleportation, quantum secret sharing \cite{HCLRL12}, and quantum error-correcting codes \cite{Sc04}. 
In general, the construction of AME states for an arbitrary number of parties and local dimensions is an open problem. An updated list of existing AME states can be found in \cite{HW_table,HESG18}. 
Interestingly, such states do not exist for a system composed of four qubits \cite{HS00}. 
Furthermore, for four subsystems
with local dimension $d$, it is possible to construct an AME state based on a classical design of two orthogonal Latin squares of size $d$.
However, this construction does not work for $d=2$
as there are no orthogonal Latin squares of size two \cite{RDK91,JCD01}.

Since the times of Euler \cite{Euler36} and Tarry \cite{GastonTarry}
it is known that two orthogonal Latin squares of order $d=6$ 
do not exist. Therefore, it was not known whether
AME states for four subsystems do exist for $d=6$ \cite{HRZ22}.
Our very recent result \cite{RBBRLZ22} can thus be considered
as a quantum solution to a classically impossible problem -- 
we constructed an explicit analytic example of a pair of
{\sl quantum orthogonal Latin squares} of size six.

The main aim of this work is to interpret and explain the solution in a possibly accessible way.
This paper is organized as follows. 
In Section II, we recall some properties of classical orthogonal Latin squares. 
Section III summons the corresponding quantum object --
quantum Latin squares, originally proposed by 
Musto and Vicary \cite{MV16}. 
In Section IV, we describe 
orthogonal quantum Latin squares
and present an exemplary configuration of size $d=6$.
In particular, we show that the construction is based on splitting the $6 \times 6=36$ dimensional 
Hilbert space into $9$ subspaces of dimension $4$ each.
An explicit form of the constructed state, a short description
of quantum entanglement and generalized Bell states,
and some further technical details
are provided in~\ref{basics} -- \ref{app:PRL_BELL}.

\section{Graeco-Latin squares and other combinatorial designs}

As a simple example of a classical combinatorial design, 
we shall recall a single Latin square. 
A \textit{Latin square} is a $d \times d$ array filled with $d$ different symbols, each occurring exactly once in each row and exactly once in each column. Below, we present an example of a $3\times 3$ Latin square.
\begin{equation} \
\large
\begin{array}{|c|c|c|} \hline 
A&C&B \\ \hline
B&A&C \\ \hline
C&B&A \\ \hline
\end{array} 
\ \sim \ 
\begin{array}{|c|c|c|} \hline 
\, 0\,& \, 2\,& \, 1\, \\ \hline
1&0&2 \\ \hline
2&1&0 \\ \hline
\end{array} 
\nonumber
 \end{equation}

\noindent
The name ``Latin square'' might be traced back to the papers of Leonhard Euler, who used Latin characters as symbols. Obviously, any set of symbols can be used: above we replaced the alphabetic sequence $A, B, C$ with the integer sequence $0, 1, 2$. An inquisitive reader might see that an analogous construction can be provided for a grid of any size $d$. 
As an example, one of the constructions consists of placing the number $i+j \;(\text{mod} \;d)$ at the intersection of the $i^{\rm th}$ column with the $j^{\rm th}$ row. 

To sum up, the general construction of a Latin square of any dimension is relatively simple. Perhaps this is why Euler became interested in constructing pairs of Latin squares that would be mutually independent in some sense. Two Latin squares of the same size $d$ are said to be \textit{orthogonal} if, by superposing symbols in each place, all possible pairs of symbols are present on the grid. A pair of \textit{orthogonal Latin squares} (OLS) has traditionally been called a \textit{Graeco-Latin square}, which refers to a popular way to represent such a pair by one Greek letter and one Latin. We present an example of a square of order 3 -- note that all nine combinations of three Latin and three Greek letters are present in the grid. 

\begin{equation} \
{\Large
\begin{array}{|c|c|c|} \hline 
{\beta\; A}    & \gamma \; C &{ \alpha \; B} \\ \hline
{ \gamma\; B} & \alpha\; A & { \beta \; C}\\ \hline
{ \alpha\; C}  & \beta \; B& {\gamma \; A}\\ \hline
\end{array} 
\ \sim \ 
\begin{array}{|c|c|c|} \hline 
{ \color{red} \texttt{K} \text{\diamond} }    & \texttt{Q} \text{\spade} &  \texttt{A} \text{\club}   \\ \hline
{ \texttt{Q}\text{\club}}  & {\color{red} \texttt{A} \text{\diamond}} &  \texttt{K} \text{\spade}    \\  \hline
{ \texttt{A} \text{\spade}}  & \texttt{K} \text{\club} &  {\color{red}  \texttt{Q} \text{\diamond} }  \\  \hline
\end{array} 
\ \sim \ 
\begin{array}{|c|c|c|} \hline 
{\color{green} 1} \, {\color{blue}0}   & {\color{green} 2}\, {\color{blue}2}&  {\color{green} 0} \, {\color{blue} 1}  \\ \hline
{\color{green} 2}\, {\color{blue}1}  & {\color{green} 0} \, {\color{blue}0} &  {\color{green} 1}\,{\color{blue} 2} \\  \hline
{\color{green} 0}\, {\color{blue}2}  & {\color{green} 1}\, {\color{blue}1}&  {\color{green} 2} \,{\color{blue}0}  \\  \hline
\end{array} 
}
\label{GL3}
\end{equation}

\noindent
Greek and Latin letters might be replaced by ranks and suits of cards or simply by pairs of numbers, as indicated above. 
Euler noticed that by replacing the pair $(a, b)$ sitting at the position $(i,j)$ of a Graeco-Latin square of size $d$  with the number $X_{ij}=ad+b+1$ the matrix $X$ forms a magic square, so the sums of entries in each row and each column are equal. 
It is easy to check that in the simplest case $d=3$ presented above this sum equals $15$. 

Constructing orthogonal Latin squares is a much harder problem than constructing a single Latin square. The problem can already be encountered in dimension $d=2$. Indeed, the desired arrangement of four pairs of Graeco-Latin characters (equivalently pairs of numbers or playing cards) is simply impossible on a $2\times 2$ grid. As we have already seen, it is possible to construct such a combinatorial design for dimension $d=3$. In fact, similar constructions can be found for any odd dimension $d$. Indeed, one can place a pair of numbers $(i+j, i+2j) \;(\text{mod} \;d)$ at the intersection of the $i^{\rm th}$ column with the $j^{\rm th}$ row. One can check that, in each column and each row of such grid all numbers: $0,\ldots,d-1$ appear on both positions, furthermore, all pairs of numbers appear on the grid. With a little more effort, the construction of orthogonal Latin squares might be given for any dimension $d$ which is a multiple of four \cite{RDK91}. In particular, for $d=4$, it takes the form of an old puzzle involving a standard deck of cards: arrange all aces, kings, queens, and jacks in a $4 \times 4$ grid such that each row and each column contains all four suits as well as one of each face value. Such a construction predate Euler and was published by Jacques Ozanam in 1725.

\begin{equation} \
{\Large
\begin{array}{|c|c|c|c|} \hline 
 \texttt{A} \text{\spade}    & { \color{red} \texttt{K} \text{\heart} }&
{ \color{red} \texttt{Q} \text{\diamond} }&  \texttt{J} \text{\club}   
\\ \hline
\texttt{Q} \text{\club}    & { \color{red} \texttt{J} \text{\diamond} }&
{ \color{red} \texttt{A} \text{\heart} }&  \texttt{K} \text{\spade}   
\\ \hline
{ \color{red} \texttt{J} \text{\heart} }&  \texttt{Q} \text{\spade}  &
\texttt{K} \text{\club}    & { \color{red} \texttt{A}\text{\diamond} }  
\\ \hline
{ \color{red} \texttt{K} \text{\diamond} }&  \texttt{A} \text{\club}  &
\texttt{J} \text{\spade}    & { \color{red} \texttt{Q}\text{\heart} }  
\\ \hline
\end{array} 
\ \sim \ 
\begin{array}{|c|c|c|c|} \hline 
 \alpha \; A    &  \beta  \; B & \gamma  \; C &  \delta \; D 
\\ \hline
\gamma \; D    & \delta \;  C & \alpha \; B &  \beta \; A   
\\ \hline
\delta \; B &  \gamma\; A  &\beta \; D   &  \alpha \; C  
\\ \hline
\beta\; C &  \alpha\; D  &\delta\; A    & \gamma\; B 
\\ \hline
\end{array} 
}
\label{GL4}
 \end{equation}

Let us summarize the rules defining a Graeco-Latin square of order $d$,
also written as OLS(d), which consists of $d^2$ symbols,
each described by its color and rank:
\medskip

{\bf 
A) all $d^2$ symbols are different,
\smallskip

B) all $d$ symbols in each row are of different colors and different ranks,
\smallskip
 
C) all $d$ symbols in each column are of different colors and different ranks.
}

\medskip
Any such solution can be written as a table of length $d^2$,
containing two numbers encoding the position in the square,
the color, and the rank. We present such a table for the case $d=3$
corresponding to the last form in pattern  (\ref{GL3}),
and label the rows and columns by indices running from $0$ to $d-1$,

\begin{equation} \
\small
\begin{array}{||c|c|c|c||} \hline \hline
    {\rm row}  & {\rm  column}  &  {\rm rank} & {\rm color}   \\ \hline
    0     &      0       & {\color{green} 1}       & {\color{blue} 0}    \\ 
    0     &      1       & {\color{green} 2}       & {\color{blue} 2}    \\ 
    0     &      2       & {\color{green} 0}       & {\color{blue} 1}    \\  \hline
    1     &      0       & {\color{green} 2}       & {\color{blue} 1}    \\ 
    1     &      1       & {\color{green} 0}       & {\color{blue} 0}    \\ 
    1     &      2       & {\color{green} 1}       & {\color{blue} 2}    \\ \hline
    2     &      0       & {\color{green} 0}       & {\color{blue} 2}    \\ 
    2     &      1       & {\color{green} 1}       & {\color{blue} 1}    \\ 
    2     &      2       & {\color{green} 2}       & {\color{blue} 0}    \\ 
   \hline \hline
\end{array} 
\label{GL5}
 \end{equation}
Using condition {\bf A}, namely the uniqueness of the symbols,
given the {\bf\color{green} rank}  and the  {\bf\color{blue} color} of a symbol 
 we can establish its position in the square, i.e.\ its row and column. 
 Conditions {\bf B}--{\bf C} imply that by knowing the {\bf column} and the  {\bf\color{blue} color}
 of a symbol we can uniquely find its {\bf\color{green} rank} and the {\bf row} in which it is located.
Similarly, once the rank (or color) and row (or column) of a symbol are given,
 we can directly infer the two other missing features.  Hence, all three conditions {\bf A}--{\bf C} are equivalent to a more general one.

\medskip 
 {\bf 
 (*) An OLS($d$) can be represented by a table consisting of $d^2$ rows, each with four numbers $(i,j,k,\ell)$ as shown in Eq.~(\ref{GL5}).
 Then, there exist three invertible functions, which map a pair of two numbers connected with any two columns into the other two: $(k,\ell)=F_1(i,j)$ and  $(j,\ell)=F_2(i,k)$ and  $(k,j)=F_3(i,\ell)$.
}
\medskip 
 
It is easy to see that in the classical case analyzed so far
 these three invertible functions 
 form permutations of size $d^2$,
 which specify, e.g.\ which symbol should be placed in which position of the Graeco-Latin square. 
As with any permutation matrix, $P$ is by construction orthogonal, its inverse is directly given by the transposition, $P^{-1}=P^T$.
 
The last array in Eq.~\eqref{GL3} encodes the permutation matrix $P_9$ of order $9$ below in the following way: for each pair $({\color{green}a},{\color{blue}b})$, the position of unity in a given column of $P_9$ is equal to $3{\color{green}a} + {\color{blue}b} + 1$. In a similar way, one can see the correspondence to the card-like version of $P_9$, where $(\texttt{A},\texttt{K},\texttt{Q})\leftrightarrow ({\color{green}0},{\color{green}1},{\color{green}2})$ and $({\color{red}\text{\diamond}},\text{\club},\text{\spade})\leftrightarrow ({\color{blue}0},{\color{blue}1},{\color{blue}2})$,
 \begin{equation}
 {\small
\label{P9}
\left[\begin{array}{ccc|ccc|ccc}
0&0&0&0&{ \color{red} \texttt{A}\text{\diamond}  }&0&0&0&0\\
0&0&\texttt{A} \text{\club}&0&0&0&0&0&0\\
0&0&0&0&0&0&\texttt{A} \text{\spade}&0&0\\
\hline
{ \color{red} \texttt{K} \text{\diamond} }&0&0&0&0&0&0&0&0\\
0&0&0&0&0&0&0&\texttt{K} \text{\club}&0\\
0&0&0&0&0&\texttt{K} \text{\spade}&0&0&0\\
\hline
0&0&0&0&0&0&0&0&{ \color{red} \texttt{Q} \text{\diamond} }\\
0&0&0&\texttt{Q} \text{\club}&0&0&0&0&0\\
0&\texttt{Q} \text{\spade}&0&0&0&0&0&0&0
\end{array}\right]
=
\left[\begin{array}{ccc|ccc|ccc}
0&0&0&0&1&0&0&0&0\\
0&0&1&0&0&0&0&0&0\\
0&0&0&0&0&0&1&0&0\\
\hline
1&0&0&0&0&0&0&0&0\\
0&0&0&0&0&0&0&1&0\\
0&0&0&0&0&1&0&0&0\\
\hline
0&0&0&0&0&0&0&0&1\\
0&0&0&1&0&0&0&0&0\\
0&1&0&0&0&0&0&0&0\\
\end{array}\right] =
P_9.}
\end{equation}
This matrix provides a simple recipe, in which place of the Latin square 
a given card has to be put. 
To assure that no color and rank repeats in each row and column of the square the non-zero entries of the matrix $P_9$ satisfy the rules of a strong Sudoku: in each row, column, and $3\times 3$ block there is a single entry equal to $1$. 
Furthermore, all the
locations of these entries in each block are different.
\medskip

As already mentioned, there does not exist an OLS of dimension $d=2$,
as locating ace and king of spades at the diagonal of the square of size two
excludes placing ace and king of hearts. 
 At the same time, we presented such construction for dimensions $d=3,4$, and $5$. The next open case is thus $d=6$, which turned out to be special. This problem was circulating in St. Petersburg in the late 1700s and, according to folklore, Catherine the Great asked Euler to solve it for a military parade. 
 This question is known as the thirty-six officers\footnote{It is sad to note that these Russian officers
  recently left their parade ground in Saint Petersburg, where they belong,
    and went a thousand miles South...
However, explicit analytical results described in this work 
strongly suggest that the officers might eventually suffer a transition
into a highly {\sl entangled state} \cite{HHHH09,WWHKT18}.
    }
problem and was introduced by Euler in the following way \cite{Euler36}. 

\hfill\begin{minipage}{\dimexpr\textwidth-3cm}
\vspace{0.5cm}
{\sl Six different regiments have six officers, each one belonging to different ranks. 
Can these 36 officers be arranged in a square formation so that each row and column contains one officer of each rank and one of each regiment?} 
\vspace{0.2cm}

-- Leonhard Euler
\vspace{0.5cm}
\end{minipage}

\begin{figure}[ht!]
\center
\includegraphics[width=1.9in]{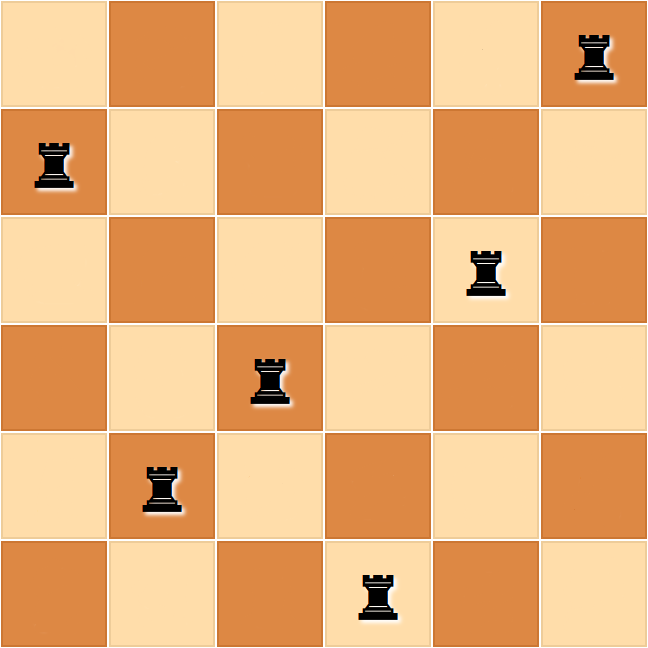} \hskip 0.6cm
\caption{A chessboard of order six with
six black rooks placed in such a way 
that no two rooks attack each other.}
\label{fig:ROOK_ONLY}
\end{figure}

\begin{figure}[ht!]
\center
a)
\includegraphics[width=2.3in]{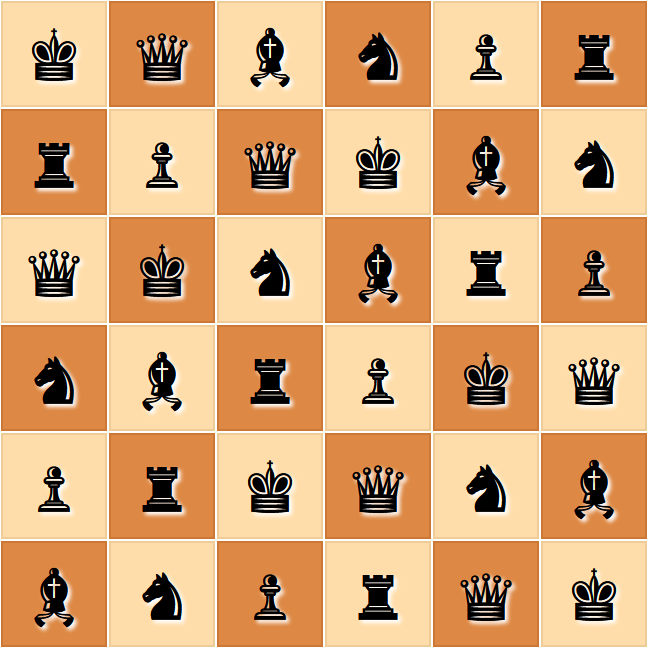} \hskip 0.6cm
b)
\includegraphics[width=2.60in]{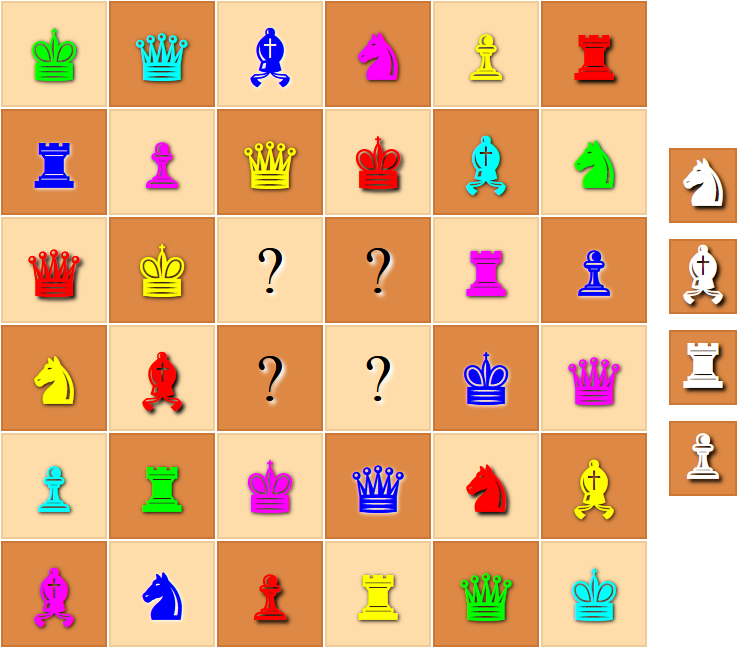}
\caption{a) A single Latin square of order six composed of $36$ black chess figures: none of the six rooks attacks any other rook (assuming that other chess pieces are removed from the board), and the same property holds also for other figures if they were replaced by rooks -- see Fig.~\ref{fig:ROOK_ONLY}.
To construct a Graeco--Latin square we have to paint the figures into six colors.
b) The problem of 36 officers of Euler in a chess setup:
check if you can place the remaining $4$ figures, two of them in
{\bf\color{cyan}cyan} and two in  {\bf\color{green}green},
on the chessboard so that it contains different figures of different colors and additionally there is no repetition of figures or colors in any row and column. Even if you do not succeed in finding an exact solution
to the problem, you will obtain a fair approximation to it -- a permutation matrix analogous to the one investigated in Ref.~\cite{CGSS05}.
}
\label{fig:d=6}
\end{figure}

To start searching for a solution
consider a simple chess problem:
Place six rooks on a
chessboard of size six, 
in such a way that no figure attacks any other.
An exemplary solution is shown in Fig \ref{fig:ROOK_ONLY}.
Let us now take six pieces of five other figures
and place them onto the board 
in an analogous way,
so that the figures of the same kind
do not attack each other, 
assuming the rook's rules
and that all other figures are not there.
Such a design, shown in Fig.~\ref{fig:d=6}a,
forms a Latin square of order six.

\noindent

To look for Graeco-Latin squares of this size
we need to color all the figures into six colors
and try to arrange them in such a way
that all colors in each row and each column
are different. 
As Euler observed, such an arrangement does not exist
-- see Fig.~\ref{fig:d=6}b -- but he was not able to prove it rigorously.
A proper proof was established only 121 years later, in a rather lengthy manner by exhaustion, by Gaston Tarry \cite{GastonTarry},
who analyzed 9408 separate cases.
A three-page short proof of the nonexistence of
a pair of orthogonal Latin squares of order six
given in 1984 by Stinson \cite{Stin84} was based on
modern results from finite vector spaces and graph theory.

On one hand, Euler was aware of methods for constructing Graeco-Latin squares in odd dimensions and a multiple of 4. On the other hand, he observed that Graeco-Latin squares of order 2 do not exist, and was unable to construct an arrangement of order 6. 
This resulted in one of his famous conjectures -- 
see \cite{KS06}.

\hfill\begin{minipage}{\dimexpr\textwidth-3cm}
\vspace{0.5cm}
There is no Graeco-Latin square  of any oddly even order $d$, i.e.\ $d \equiv 2 $ (mod $4$). 
\vspace{0.2cm}

-- \emph{Euler's Graeco-Latin square conjecture}
\vspace{0.5cm}
\end{minipage}

Euler's conjecture remained unsolved until 1959 when Raj Chandra Bose and Sharadchandra Shankar Shrikhande constructed a counterexample of order $d=22$ \cite{BS59}. 
In the same year, Ernest Tilden Parker found a lower-dimensional counterexample of order $d=10$ using a computer search, which was one of the earliest combinatorics problems solved by means of a digital computer \cite{JCD01}.
Later on, this serious mathematical result was popularized in a renowned
novel by Perec \cite{Pe78},
in which the $d=10$ Graeco-Latin square plays a crucial role.

 In April 1959, all three mathematicians: Parker, Bose, and Shrikhande, together generalized the results and proved Euler's conjecture to be false for all $d \geq 10$ by constructing related arrangements \cite{BSP60}.
The result of these ``Euler's spoilers" as they were dubbed, completed the answer to the question: 

\hfill\begin{minipage}{\dimexpr\textwidth-3cm}
\vspace{0.5cm}
For which number $d$ there exists a Graeco-Latin square of order $d$?
\vspace{0.2cm}

\emph{Answer}: It exists for all numbers $d$ except two and six. 
\vspace{0.5cm}
\end{minipage}

It is worth mentioning that Euler was not only interested in pairs of orthogonal Latin squares, which we called after him Graeco-Latin squares, but also by tuples of such squares which are pairwise orthogonal. A set of $n$ Latin squares of the same order $d$ in which all pairs are orthogonal is called a set of \textit{mutually orthogonal Latin squares} (MOLS). Below, we present an example of three mutually orthogonal Latin squares of order $4$, where, in addition to Greek and Latin letters in (\ref{GL4}), we include here also four Hebrew: $\aleph, \beth,\gimel,\daleth$.
\begin{equation} \
{\large
\begin{array}{|c|c|c|c|} \hline 
 \alpha \; A \; \aleph  &  \beta  \; B\; \beth & \gamma  \; C\; \gimel &  \delta \; D \;\daleth
\\ \hline
\gamma \; D \; \beth  & \delta \;  C \; \aleph & \alpha \; B \;\daleth &  \beta \; A   \;\gimel
\\ \hline
\delta \; B \;\gimel &  \gamma\; A  \;\daleth &\beta \; D \; \aleph &  \alpha \; C  \; \beth
\\ \hline
\beta\; C \;\daleth&  \alpha\; D \; \gimel &\delta\; A  \; \beth & \gamma\; B \; \aleph
\\ \hline
\end{array} 
\ \sim \ 
\begin{array}{|c|c|c|c|} \hline 
0 \; {\color{blue} 0} \; {\color{red} 0} &  1 \; {\color{blue} 1}\; {\color{red} 1} & 2  \; {\color{blue} 2}\; {\color{red} 2} &  3 \; {\color{blue} 3} \;{\color{red} 3}

\\ \hline

2 \; {\color{blue} 3} \; {\color{red} 1}  & 3 \;  {\color{blue} 2} \; {\color{red} 0} & 1 \; {\color{blue} 1} \;{\color{red} 3} &  2 \; {\color{blue} 0}   \;{\color{red} 2}

\\ \hline

3 \; {\color{blue} 1} \;{\color{red} 2} &  2\; {\color{blue} 0}  \;{\color{red} 3} &2 \; {\color{blue} 3} \; {\color{red} 0} &  1 \; {\color{blue} 2}  \; {\color{red} 1}

\\ \hline

2\; {\color{blue} 2} \;{\color{red} 3}&  1\; {\color{blue} 3}\; {\color{red} 2} &3\; {\color{blue} 0}  \; {\color{red} 1} & 2\; {\color{blue} 1} \; {\color{red} 0}

\\ \hline
\end{array} 
}
\label{GLH4}
 \end{equation}

\noindent
As we have already seen, constructing even a single pair of OLS is not a simple task in some dimensions 
$d$. 
Up to this date, the following general question regarding MOLS remains open. 

\hfill\begin{minipage}{\dimexpr\textwidth-3cm}
\vspace{0.5cm}
What is the maximum number of mutually orthogonal Latin squares of a given order $d$?
\vspace{0.5cm}
\end{minipage}

\noindent
There are some partial results regarding answers to the above question. Firstly, it might be observed that for each dimension $d$ there are at most $d-1$ MOLS \cite{JCD01}.
Therefore the Graeco-Latin square (\ref{GL3}) cannot be extended into three MOLS of order three, similarly, the set of three MOLS of order four (\ref{GLH4}) cannot be extended further.
Such a maximal set of $d-1$ MOLS is known to exist for all prime and prime power dimensions $d$. Furthermore, the aforementioned results concerning Graeco-Latin squares show that in any dimension apart from $d=2,6$ there are at least two MOLS. While in the case $d=2$, the single existing Latin square saturates the upper bound, $d-1=1$, the dimension $d= 2\times 3=6$ is entirely different, as the upper bound implies that in this case there are no more than $d-1=5$ MOLS.

\section{Quantum Latin squares}

Looking for a classical combinatorial design we analyze constellations 
formed of various objects belonging to a given discrete set of numbers, letters,
cards, or chess figures. 
In quantum theory, a state is described by 
a  vector $|\psi\rangle$ from a complex $d$-dimensional Hilbert space ${\cal H}_d$.
We shall assume that such vectors are normalized,
$||\psi||^2=\langle \psi|\psi\rangle=1$,
and that two vectors that differ by a complex phase are identified to be the same, according to an equivalence class given by $|\psi\rangle \sim e^{i \alpha} |\psi\rangle$.
Then, the set of quantum states is continuous,
as any two states can be connected by a unitary transformation,
$|\psi\rangle = U |\phi\rangle$.
Two quantum states   $|\psi\rangle$ and $|\phi\rangle$ are distinguishable 
if and only if they are orthogonal, 
$\langle \psi|\phi\rangle=0$. A set of $d$ different classical objects
labeled by $i=0,1\dots, d-1$ corresponds to an orthogonal  
basis $|i\rangle \in {\cal H}_d$ of distinguishable states, $\langle i|j \rangle=\delta_{ij}$.

Apart from the classical states like $|0\rangle$ and $|1\rangle$
one can consider their arbitrary superpositions,
say $|\Psi_{\theta,\chi}\rangle=\cos\theta |0\rangle + e^{i \chi} \sin\theta |1\rangle$.
In the case of bipartite systems consisting of two subsystems $A$ and $B$, one uses a composed Hilbert space defined by the tensor product,
 ${\cal H}_{AB} ={\cal H}_A \otimes {\cal H}_B$.
 A separable state has a {\sl product form}, 
$|\psi_{AB}\rangle =|\phi_A\rangle \otimes |\phi_B\rangle$,
for brevity written as $|\phi_A,\phi_B\rangle$ or $|\phi_A\, \phi_B\rangle$.
All other (non-product) states are called {\sl entangled} \cite{BZ17}. 
A key example is given by the celebrated {\sl Bell state},
$|\phi_+\rangle :=(|00\rangle +|11\rangle)/\sqrt{2}$ -- for details, see~\ref{basics}.

A rule of thumb says that any good notion can be quantized.
Following this reasoning, recently Musto and Vicary defined quantum Latin square (QLS)~\cite{MV16}.
A QLS of size $d$ consists of $d^2$ quantum states
$|\psi_{ij}\rangle \in {\cal H}_d$ for $i,j=0,\dots, d-1$,
such that each row and each column forms an orthogonal basis.
Note that the classical condition: all objects in each row and each
column are {\sl different}, is now replaced by its natural quantum analog:
all quantum states in each row and column are {\sl orthogonal}.

\begin{figure}[ht!]
\center
\includegraphics[height=2.5cm]{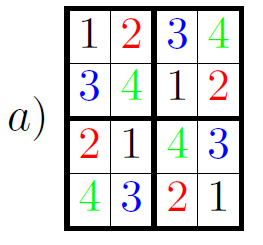}
\includegraphics[height=2.5cm]{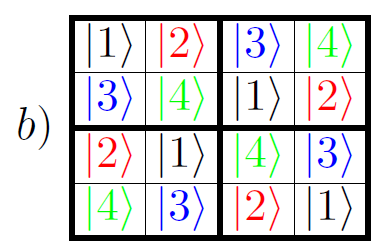}
\includegraphics[height=2.5cm]{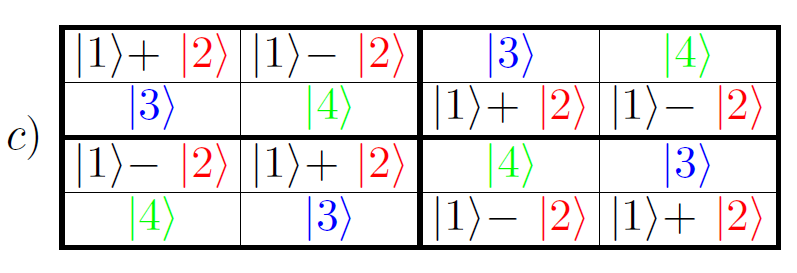}
\includegraphics[height=2.7cm]{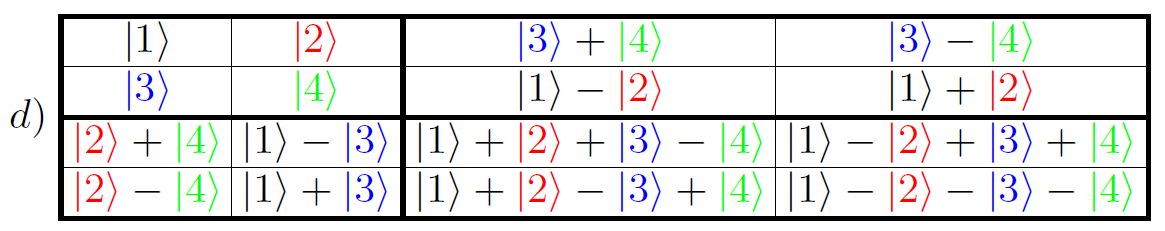}
\caption{Examples of Latin squares of size $4$ with the Sudoku property  --
   each square of size $2$ contains  different classical elements 
   or orthogonal quantum states: 
 a) classical Latin square related to movement of chess knight;
 b) its direct quantum analog;
 c) apparently quantum Latin square with $C=4=d$ different states which can be rotated 
 to the computational basis;
d) genuinely quantum Latin square with $C=16$ different states forming 
$4 \times 3=12$ different orthogonal bases in ${\cal H}_4$.
 For legibility, quantum states are not normalized 
 and we use here a notation where the kets are numbered from $1$ do $d$.}
\label{fig:sudoku}
\end{figure}

Fig.~\ref{fig:sudoku}a shows a classical Latin square of order $d=4$,
while  Fig.~\ref{fig:sudoku}b presents its direct quantum analog.
The labels run here from $1$ to $d=4$.
It is clear that all orthogonality rules are satisfied,
but this object does not differ much from its classical ancestor.
Observe that Fig.~\ref{fig:sudoku}c looks "more quantum"
as it contains superposition states,
$|\pm\rangle =|1\rangle \pm |2\rangle$.
However, this design is only {\sl apparently quantum},
as it can be converted to a classical one by a unitary rotation.
 To characterize the degree of quantumness it is sufficient
 to count the number of different states in the pattern 
 -- this number is called the {\sl cardinality} $C$ of a QLS.
 Any design of size $d$ with cardinality $C=d$ 
 is apparently quantum, as it is equivalent with respect to
 a unitary rotation to a classical design. Hence all orthogonal bases
corresponding to various rows and columns of the pattern are equal. 
  On the other hand,
 if $C>d$ this is no longer the case and such a design is called
 {\sl genuinely quantum}.
 
It is possible to show  \cite{PWRBZ21}  that in dimensions $d=2$ and $d=3$ 
all quantum Latin squares satisfy the 
relation $C=d$, so they are apparently quantum.
For $d=4$ there exist QLS with cardinality greater than four,
which are genuinely quantum and are not equivalent  to any classical solution. 
 Fig.~\ref{fig:sudoku}d  shows the quantum pattern with the
 maximal cardinality, $C=d^2=16$.
 It satisfies the rules of a quantum Sudoku \cite{NP21},
 so it offers a constellation of  $3 \times 4=12$ different 
 orthogonal measurements in ${\cal H}_4$.
 Further examples of QLS(d) with maximal cardinality $C=d^2$
 were constructed in \cite{PWRBZ21} for an arbitrary dimension $d$.

\section{Quantum orthogonal Latin squares}

Quantization of a classical Graeco--Latin square of size $d$
seems to be straightforward.
In each field of the square it is enough to place a product state 
$|ik\rangle = |i\rangle \otimes |k\rangle$,
where  $i,k=0,\dots, d-1$. 
In such a way 
all $d^2$ entries of a square form a product basis in 
${\cal H}_d \otimes {\cal H}_d$.
 This construction,
   explicitly visualized for $d=3$ in Eq.~(\ref{GL7}), 
 corresponds directly to the classical  design shown in Eq.~(\ref{GL3}).

\begin{equation} \
{\Large  
\begin{array} {|c|c|c|} \hline 
 {\color{green} |1\rangle}  \otimes {\color{blue}  |0\rangle}  
 & {\color{green} |2\rangle}   \otimes {\color{blue}  |2\rangle}  
 & {\color{green} |0\rangle}  \otimes {\color{blue} 
     | 1 \rangle}
    \\ \hline
{\color{green} |2 \rangle} \otimes  {\color{blue}|1\rangle}  
& {\color{green} |0\rangle} \otimes {\color{blue}|0\rangle} 
&  {\color{green} |1\rangle}\otimes {\color{blue} |2\rangle} 
\\  \hline
{\color{green} |0\rangle } \otimes {\color{blue} |2\rangle} 
 & {\color{green} |1\rangle}\otimes {\color{blue}|1\rangle}
 &  {\color{green} |2\rangle } \otimes {\color{blue}|0\rangle} 
  \\  \hline
\end{array} 
}
\label{GL7}
 \end{equation}

More formally, a pair of  quantum  orthogonal Latin squares (QOLS) 
of size $d$
can be defined as a set of $d^2$ bipartite states,
$|\psi_{ij}\rangle \in {\cal H}_A \otimes {\cal H}_B$
which form a square,
\begin{equation} \
{\large 
{\rm QOLS} \; :=  \;
\begin{array} {|c c c|} \hline 
|\psi_{11}\rangle & \dots & |\psi_{1d}\rangle \\
      \dots            & \dots  &    \dots     \\
|\psi_{d1}\rangle & \dots & |\psi_{dd}\rangle \\
\hline
\end{array} \; ,
}
\label{QOLS2}
 \end{equation}
and satisfy certain constraints analogous to the classical conditions {\bf A}--{\bf C} above. 
As the bipartite states 
$|\psi_{ij}\rangle$, in general, can be entangled, 
it is not always possible to split 
such a pair of QOLS into two separate
orthogonal quantum Latin squares.

Condition {\bf A}, imposing that all the objects of the classical design are different,
 is easy to quantize:

\medskip
{\bf A') All $d^2$ states are mutually orthogonal and they form a basis},
\begin{equation}
\label{a_prim}
\langle \psi_{ij}|\psi_{k\ell}\rangle \; =\;  \delta_{ik} \delta_{j \ell}.
\end{equation}

\medskip

Two further conditions {\bf B}--{\bf C} are more involved.
Since in each row (or column) we
analyze {\sl only} ranks or {\sl only} colors
of the chess pieces in the design, 
it is natural to expect that
the corresponding quantum constraints will
deal with the {\sl partial trace} (see~\ref{basics})
of a symmetric combination of projectors.
Investigating the QOLS shown in (\ref{GL7}) we realize 
that the  superposition of bipartite states along the second  column 
forms a maximally entangled state,
$|\gamma_2\rangle = (|22\rangle + |00\rangle +|11\rangle)/\sqrt{3}$.
Thus, the partial trace over the second subsystem $B$ (or the first subsystem $A$)
is maximally mixed,
${\rm Tr}_B |\gamma_2\rangle \langle \gamma_2|\propto{\mathbbm I}_3$, as all elements in this column of the design are different.
It is easy to see that the superpositions of three states in each column and each
row have the same property as they are locally equivalent to the generalized Bell state,
 $|\gamma_2\rangle\in {\cal H}_3 \otimes {\cal H}_3$.
Such a condition
was initially advocated in \cite{GRMZ18} to become a part of a possible
definition of QOLS. However, it occurred \cite{MV19,Ri20}
that these conditions, satisfied for superposition of product states, 
are not sufficient to assure the required properties of an AME
state of four subsystems, with $d$ levels reach. 
Necessary and sufficient conditions can be formulated 
in terms of the partial trace of sums of projectors in each row and column \cite{GRM21},

\medskip
{\bf B') All rows of the square satisfy the conditions}
\begin{equation}
\label{b_prim}
{\rm Tr}_B \Bigl( \sum_{k=0}^{d-1}  |\psi_{ik}\rangle \langle \psi_{jk}|  \Bigr) \; =\; \delta_{ij}
{\mathbbm I}_d\;.
\end{equation}
Analogous to dealing only with the colors of pieces in each row and forgetting their ranks,
 and dual conditions for the other partial trace ${\rm Tr}_A$ analogous to the requirement that all ranks in each row of the classical design are different.
 
\medskip

{\bf C') All columns of the square satisfy the partial trace conditions
\begin{equation}
\label{c_prim}
{\rm Tr}_B   \Bigl( \sum_{k=0}^{d-1}  |\psi_{ki}\rangle \langle \psi_{kj}| \Bigr) \;= \; \delta_{ij} 
{\mathbbm I}_d \; ,
\end{equation}
 \hskip 1.3cm  and dual conditions for ${\rm Tr}_A$.
\medskip
}

These general conditions can be understood in analogy to their classical counterparts.
Namely, if $i=j$, conditions {\bf B'} and {\bf C'} with the identity mean that each symbol occurs exactly once -- the mixing of features (rank, suits, etc.) is perfect.
For $i\neq j$, the vanishing value of the partial trace is assured by the fact that
while taking two distinct columns, the property of being a Latin square means that no feature is repeated.

It is possible to show~\cite{GRM21}
that the above conditions,
easy to verify and visualize,
are equivalent to those used in the former paper \cite{RBBRLZ22} 
and in earlier works on AME states \cite{Sc04,BP+07,FFPP08, HCLRL12}.
The above conditions are invariant with respect to 
local unitary operations $U_A \otimes U_B$,
so they are by construction satisfied 
by apparently orthogonal quantum Latin squares
obtained by a local unitary transformation of any classical solution. 
See~\ref{app:visual_ptrace} for a visual explanation of partial tracing.

Another option to analyze a QOLS consists in representing
the square as a table with $d^2$ rows
and $4$ columns analogous to (\ref{GL5}).
However, this scheme is not yet defined well enough,
as the states can be entangled. 
Placing into the first field of 
(\ref{GL4}), for instance, 
$|\psi_{11}\rangle =  (|\texttt{A} \text{\spade} \rangle + |{\color{red} \texttt{K} \text{\heart}}\rangle)/\sqrt{2}$,
we see that neither the color nor the rank of this
object is well specified.
However, it is possible to modify the classical condition ({\bf *})
introducing the corresponding four-party AME state \cite{RBBRLZ22}
\begin{equation}
\label{AME4}
|\Psi_{ABCD}\rangle = \frac{1}{d} \sum_{i,j=0}^{d-1}  |i\rangle_A |j\rangle_B |\psi_{ij}\rangle_{CD}
= \frac{1}{d} \sum_{i,j,k,\ell=0}^{d-1} T_{ijk\ell}   |i\rangle_A |j\rangle_B |k\rangle_C |\ell\rangle_{D}
\end{equation}
This form of representing the square consisting of $d^2$ bi-partite states $|\psi_{ij}\rangle$
allows one to formulate a single general condition analogous to the classical one ({\bf *}), 

\medskip
{\bf (**) 
 Representing a quantum design (\ref{QOLS2}) by a tensor $T_{ijk\ell}$  
 with four indices, each running from $0$ to $d-1$,  introduced in  (\ref{AME4}),
 we say that it forms an 
 orthogonal quantum Latin square if all three matrices 
 obtained by various pairing of two indices, 
 $X_{\mu,\nu}=T_{ij,k\ell}$,
 $Y_{\mu,\nu}=T_{ik,j\ell}$ and
 $Z_{\mu,\nu}=T_{i\ell,kj}$, form unitary matrices of size $d^2$.
}

\medskip

Observe that in the classical case these matrices form three orthogonal
permutation matrices of size $d^2$, mentioned in condition ({\bf *}), exemplified by  Eq.~(\ref{P9})
and its reorderings for $d=3$.
Given a matrix $X$ with entries $X_{\mu,\nu}=X_{ij,k\ell}$
the matrix $Y$ with entries $Y_{ij,k\ell}=X_{ik,j\ell}$, 
written, $Y=X^{\rm \Gamma}$, is called partial transpose of $X$,
while matrix $Z$ with entries $Z_{ij,k\ell}=X_{i\ell, jk}$,
written, $Z=X^{\rm R}$, is called reshuffling of $X$.
More details concerning this notation can be found 
in the Supplementary Material of the previous work \cite{RBBRLZ22}. 
A tensor $T$ satisfying such conditions is called {perfect}	\cite{PYHP15},
while a unitary matrix $U$ of size $d^2$, 
which remains unitary under two corresponding exchanges of indices is called
{$2$-unitary} \cite{GALR15}.

\begin{figure}[ht!]
\center
\includegraphics[height=3cm]{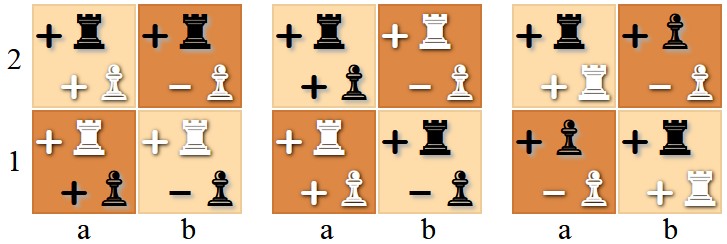}
\caption{Exemplary configurations of four two-qubits entangled
states represented on a chessboard of size $2$.
None of the above configurations satisfies the necessary conditions
{\bf A'}--{\bf C'}).
In the array on the left, the first and second columns do not satisfy \textbf{C'} condition while tracing over colors. 
The proof that the other two arrangements do not satisfy some conditions is left as an exercise to the reader.
}
\label{fig:AME42}
\end{figure}

\bigskip

Before presenting a solution to a QOLS for $d=6$
let us return first to the simpler case $d=2$.
Fig.~\ref{fig:AME42}
shows potential designs, composed of a basis of four
orthogonal Bell states.
It is possible to demonstrate that 
the presented structures do not satisfy all necessary
 conditions {\bf A'}--{\bf C'}.
 The  fact that there are no QOLS of order $d=2$
  is equivalent to 
  the seminal result of Higuchi and Sudbery, 
who proved that there are no AME states for a 
system composed of four qubits \cite{HS00}.

\section{Thirty-six entangled officers of Euler}

As there are no solutions to the classical problem of 36 officers of Euler,
we cannot look for a quantum design (\ref{QOLS2})
taking the bi-partite states $|\psi_{ij}\rangle$
 in a product form as it was done for $d=3$ in (\ref{GL7}).
 Instead, one has to allow the officers to be entangled
 and use entangled states visualized by several chess figures 
 occupying the same field of the chessboard.
 For instance, in place of putting  a classical  red king {\color{red}\kingB}
into the chess field \texttt{a1} and a blue queen { \color{blue}\queenB} into \texttt{a2},
we can locate a Bell state
$|{\color{red}\kingB}\rangle$ $+$ $|{\color{blue} \queenB}\rangle$
into the first field and an orthogonal state
$|{\color{red}\kingB}\rangle$ $-$  $|{\color{blue}\queenB}\rangle$
into the second one (for simplicity we shall now use non-normalized states).
Entangled states,
like $|{\color{blue}\kingB}\rangle$ $+$ $|{\color{red} \queenB}\rangle$,
can contain arbitrary complex phases and the number of states in a superposition can be larger,
so it is not possible to exhaust all possible configurations on a chessboard.

The original solution to the problem
was found numerically \cite{RBBRLZ22,GRM21,WB_thesis} using a Sinkhorn-like algorithm,
earlier applied in \cite{SAA2020} 
to find  dual  unitary matrices,  for which we relaxed one condition and required that only
$U$ and $U^{\rm R}$ are unitary \cite{BKP2019}. 
Later on, we have found a purely analytic form of the solution \cite{AB_thesis} that we shall now present in some detail.

\begin{figure}[ht!]
\center
\includegraphics[width=3.0in]{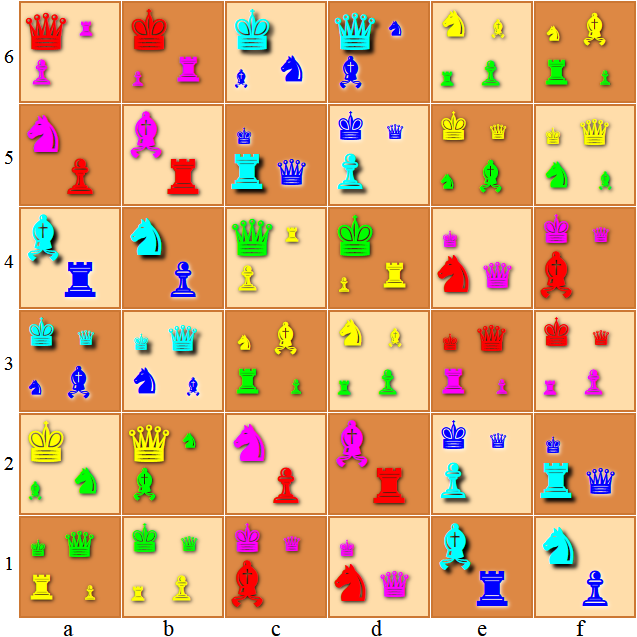}
\caption{Chess-like illustration of a solution to the QOLS(6) problem corresponding to the
matrix $U$. Identify nine squares of size two each
consisting of $4$ fields, each containing figures in only two colors.
Note that such a coarse-grained structure forms a Latin square of size $3$. 
}
\label{fig:UG_braz}
\end{figure}

A simplified version of the solution to the problem of 36 entangled officers
is visualized in Fig.~\ref{fig:UG_braz}  on a truncated chessboard of size $6 \times 6$.
Each field corresponds to one officer, or a quantum state 
$|\psi_{ij}\rangle \in {\cal H}_6 \otimes {\cal H}_6$,
or a row of  a 2-unitary matrix $U \in U(36)$.
A~chess field with two or more different figures in different colors represents an entangled state.
Interestingly, there are at most four figures in each field,
which means that the entanglement concerns only $2,3$,
or $4$ officers out of $36$.

 The size of each figure represents its relative weight in the superposition, so the fields
 with two figures of the same size
  represent Bell states. 
  The ratio of the larger to the smaller sizes in the fields 
  hosting four figures is equal to the golden mean, $b/a=(1+\sqrt{5})/2=\varphi$,
  with $a=\frac{1}{2}\sqrt{1-1/\sqrt{5}}$ and $b=\frac{1}{2}\sqrt{1+1/\sqrt{5}}$.
The third and the largest amplitude is $c=1/\sqrt{2}$ and they all fulfill the Pythagorean relation: $a^2+b^2=c^2$.

\begin{figure}[ht!]
\center
\includegraphics[width=3.2in]{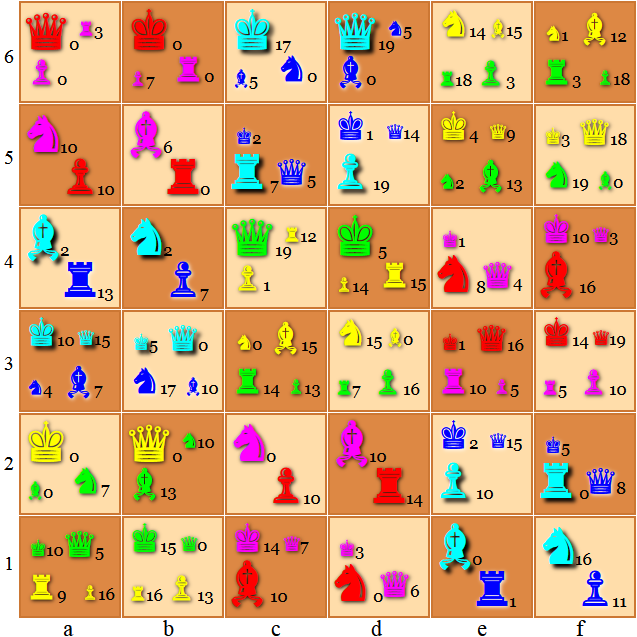}
\caption{Number $k\in\{0,1,...,19\}$ decorating each figure denotes
  the complex phase, $\exp(2 \pi i k/20)$,
  making the representation of the QOLS(6) complete.
This single figure encodes the entire solution to the problem!
See~\ref{AME_full_formulas} for a detailed explanation.
}
\label{fig:UG_phase}
\end{figure}

However, a reader with a sharp eye will immediately identify a problem:
the fields \texttt{c2} and \texttt{a5} contain a red pawn and a purple knight, so they look the same,
although they should be distinguishable!
The answer is contained in the complex phases of each state
shown explicitly in Fig.~\ref{fig:UG_phase}.
An integer  number $k$ labeling each figure 
denotes the complex phase $\omega^k$, with $\omega=\exp(2 \pi i /20)$ equal to the $20^{\rm th}$ root of unity. 
The phase $\omega$ plays an important  geometric role 
as it is related to the regular pentagon and the golden number $\varphi$, see~\cite{RBBRLZ22}.
Since the phase $k=10$ corresponds to $-1$ the chess field \texttt{c2} represents the Bell state
 $|{\color{magenta}\knightB}\rangle$ $-$ $|{ \color{red}\pawnB}\rangle$.
Note also that the other field \texttt{a5} corresponds to another, orthogonal, Bell state
equivalent to 
 $|{\color{magenta}\knightB}\rangle$ $+$ $|{ \color{red}\pawnB}\rangle$,
 so both states are perfectly distinguishable, as they should be. Similar reasoning shows
 that all $36$ states, uniquely determined by each chess field,
  are mutually orthogonal.
  
\begin{figure}[ht!]
\center
a) \! \! \includegraphics[width=2.2in]{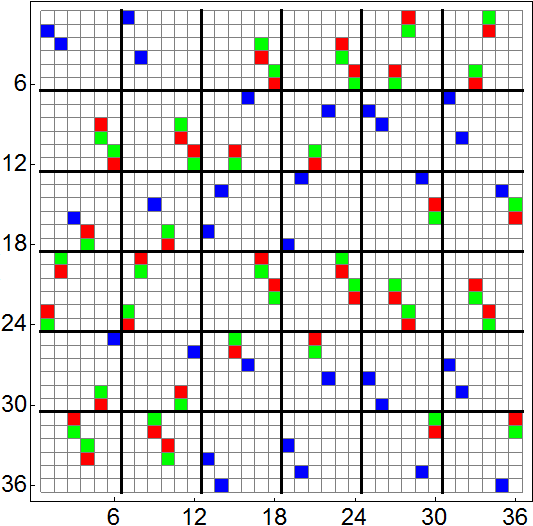} \hskip 0.5cm
b) \!\! \includegraphics[width=2.2in]{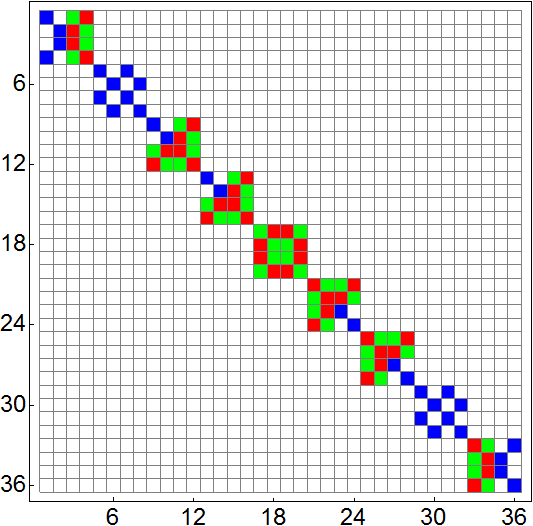}
\caption{a) The structure of the unitary matrix $U$ of size $36$
which describes the quantum design: each row represents a single officer
and it has at most $4$ non-zero entries. The colors describe different amplitudes; red for $a$, green for $b$, and blue for $c$. Note that the color
used here have nothing in common with the colors in other figures.
b) The matrix $U$ is permutation-equivalent to a block-diagonal form,
which contains a direct sum of nine unitary blocks of order four. Compare this
with~\ref{app:PRL_BELL}.
}
\label{fig:matrix_U}
\end{figure}

This implies that condition  (\ref{a_prim})
of perfect distinguishability of all entries of the designs is satisfied.
Hence, the matrix $X$ of size $36$ representing the design is unitary, so it will now be referred to as $U$. Its structure is shown in Fig.~\ref{fig:matrix_U}a,
  in which three colors represent three different amplitudes.
  Note that the matrix is sparse -- each row contains at most $4$ non-zero entries,
  as there are at most $4$ figures in each chess field.  Performing a suitable permutation 
  of size $36$ it is possible to bring the matrix  into a block diagonal form:
  as shown in Fig.~\ref{fig:matrix_U}b, it consists of nine unitary blocks of size $4$.
  In such a way entire space of size $36$ can be decomposed into nine subspaces of size four.
 
\begin{figure}[ht!]
\center
\includegraphics[width=3.1in]{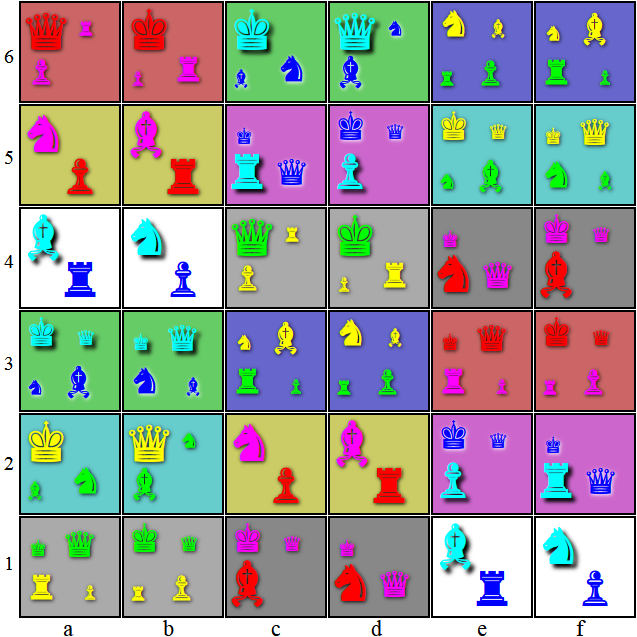}
\caption{The same chessboard as in Fig.~\ref{fig:UG_braz}
with nine background colors representing each 4-dimensional subspace
and corresponding to each block in Fig.~\ref{fig:matrix_U}b.
}
\label{fig:UG_background}
\end{figure}

One can also check that the two other necessary conditions
\eqref{b_prim} and~\eqref{c_prim}
 hold true. To check this, one can take into account the decomposition
$36=9 \times 4$, as shown in Fig.~\ref{fig:UG_background},
in which each four-dimensional base is labeled by a background color.
Note that each row and each column of the design
contains entries of three of six different  bases 
which play a role in the diagonal part, $i=j$, of conditions,  {\bf B'} and {\bf C'}.
Furthermore, checking the off-diagonal constraints for  $i\ne j$,
one observes that by choosing any pair of rows (or columns) of the
square and comparing their corresponding entries,
only two different configurations are possible.
In the first case (example: the first and the second column)
all pairs of  fields have the same color of the background, 
so the corresponding states belong to the same $4{\rm D}$ subspace.  
In the second case, (example: first and third column 
or any two rows), all pairs of squares have different colors,
so the states belong to orthogonal subspaces.

\begin{figure}[ht!]
\center
a)
\includegraphics[width=2.5in]{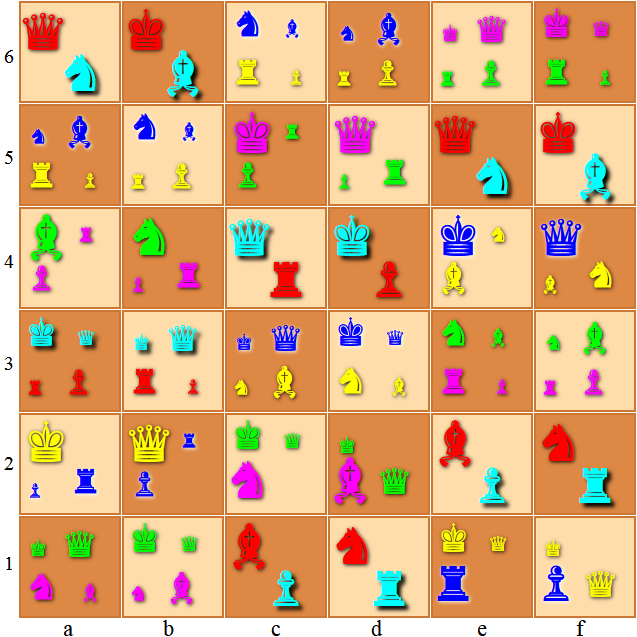} \hskip 0.6cm
b)
\includegraphics[width=2.5in]{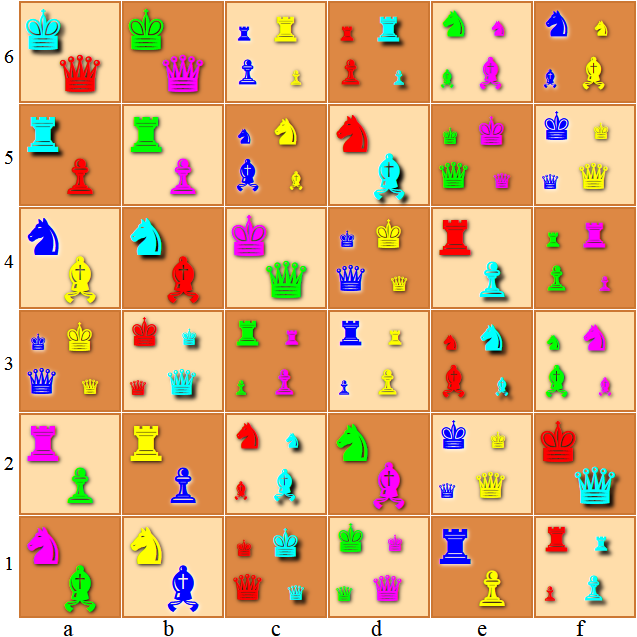}
\caption{The same QOLS(6) solution now corresponding to 
a) the reshuffled matrix $U^{\rm R}$ and b)  the partially transposed matrix $U^{\rm \Gamma}$. 
Note that after the reorderings of the matrices
 the structure of each field and colors and coupled figures do change. 
}
\label{fig:UG_UGR}
\end{figure}

As the matrix $U$ and the set of $36$ states $|\psi_{ij}\rangle$
 determined by its rows and represented in a chessboard
satisfy all conditions~\eqref{a_prim}, \eqref{b_prim}, and \eqref{c_prim},
the design forms a constructive example of QOLS(6).
It is also instructive to look at this solution analyzed from another perspective,
choosing another basis. If a matrix $U$ forms a valid solution to the problem
also reordered matrices $U^{\rm R}$ and $U^{\rm \Gamma}$ exhibit  the same property.
The corresponding two (equivalent) solutions
are shown in two chessboards presented in Fig.~\ref{fig:UG_UGR}.
Furthermore, it is possible to show that
every state out of $36$, living in a $4$--dimensional subspace, is locally equivalent to the standard, two-qubit Bell state
 $|\phi_+\rangle  =(|00\rangle +|11\rangle)/\sqrt{2}$.

Looking at the structure of the representation of $U^{\rm \Gamma}$,
shown in Fig.~\ref{fig:UG_UGR}b, we realize that 
kings are only entangled with queens or other kings, bishops with knights, and rooks with pawns. 
Using another vocabulary, all generals talk only to admirals,
colonels to majors, while captains are only entangled with lieutenants.
Furthermore, six colors split into three pairs of colors which produce a single state,
hence the entire cohort of $36$ officers splits into nine groups of size four, according to the
title of this paper.

\section{Concluding remarks}

The famous combinatorial problem of 36 officers was posed by Euler, who 
claimed in 1779 that no solution exists. 
The first paper with proof of this statement, by Tarry \cite{GastonTarry}, came only 121 years later, in 1900.
After another 121 years, we have presented a
solution to the quantum analog of the Euler problem \cite{RBBRLZ22},
in which superpositions of officers and entanglement between their ranks and their units is allowed.

This unexpected result implies constructive solutions to the related problems of the existence of 
absolutely maximally entangled states of four subsystems with six levels each, a $2$-unitary matrix $\mathcal{U}_{36}$ of size $36$ with maximal entangling power \cite{GALR15,SAA2020}
and a perfect tensor $T_{ijk \ell}$ with four indices, each running from one to six
\cite{PYHP15}.
Our results allowed us to construct original quantum error-correcting codes 
\cite{Sc04}: a pure code
usually denoted as  $(\!(4,1,3)\!)_6$, and a shortened code $(\!(3,6,2)\!)_6$, 
which allows encoding a $6$-level state into a set of three such subsystems. 
Furthermore, it implies the existence of a quantum orthogonal array \cite{GRMZ18} of strength two with six rows and four columns: two classical and two quantum, written QOA$(6,2+2,2,6)$. 

Although we worked for more than a year to understand 
the fine mathematical structure of the solution found,
it is clear that more effort is needed to provide its
 optimal  explanation, accessible to a broad audience.
 In this paper we used a chess analogy, first advocated in  \cite{Ga22}.

Observe a similarity between our solution and the games played with "quantum chess"
\cite{Daniel}, 
in which a player can position a single piece in multiple locations on the board at once!
 In both problems the quantum rules allow one to create a superposition of states corresponding to various chess pieces at different locations.

 Our solution to the quantum design of $36$ entangled officers of Euler was initially obtained by an extensive numerical search. 
 A posteriori, applying reverse engineering,  we can now
propose a general construction scheme of the unitary matrix 
$U\in U(36)$,  which encompasses the entire solution:

\medskip

i)  Generate nine copies of  Bell bases, represented by a simple sum
     of  nine copies of Bell matrix (\ref{bell}), written  $U'=\oplus_{i=1}^9 B_i$ 
     with $B_i=B$.
    
    \smallskip
     
    
    ii) Rotate each basis locally and apply a diagonal matrix of phases, $B'_i=(W_i\otimes {\tilde W}_i) B_i D_i$,
   with local unitary matrices $W_i$, ${\tilde W}_i$ and a diagonal unitary matrix $D_i$,
   to obtain a unitary block matrix $U''=\oplus_{i=1}^9 B'_i$,
    similar to the one shown in Fig.~\ref{fig:matrix_U}b.
    
    \smallskip
    
 iii) Permute the entire matrix by suitable permutation matrices $P_1$ and $P_2$   
    of size $36$ to get the form
      $U^{\Gamma}=P_1U''P_2$. 
      This scheme allows one to obtain the partially transposed matrix
  presented in Fig.~\ref{fig:UG_UGR}b,
  so to get the final matrix 
presented in Fig.~\ref{fig:matrix_U}a
  one needs to perform a partial transpose,
  $U= (U^{\Gamma})^{\Gamma}$.
  
  \medskip 
  
Unfortunately, this rather simple  prescription is not a constructive one,
as it does not specify explicitly $18$ unitary matrices $W_i$ and ${\tilde W}_i$
   of order two, diagonal matrices $D_i$,
   and additionally, 
   two permutation matrices $P_1$ and $P_2$ of size $36$.
     
\bigskip

Let us conclude this paper by posing a list of natural open questions:
\smallskip

a)  Does there exist a real solution to the problem of entangled 36 officers of Euler? 
In another language, we ask if an orthogonal matrix of size 36 with  a 2-unitary property can be found. 
In particular, one can ask if there exists a 2-unitary Hadamard matrix of size $d^2=36$?
\smallskip

b)  Does there exist a QOLS(6) not equivalent to the form presented
      with respect to local unitary transformations?
\smallskip
  
c)   Are there genuinely quantum OLS(d) for $d=3$, $4$ or $5$, 
      which are not locally equivalent to a classical solution?
\smallskip

d)  Can we change the title equation into e.g.
    $4 \times 16 = 8 \times 8$,  and find a genuinely quantum solution  
       for QOLS(8) based on $16$ Bell bases of maximally entangled two qubit states,
          or a two unitary matrix of size $64$, up to permutation equivalence to 
           a block matrix with $16$ blocks of size $4$?
                
\newpage

{\bf Acknowledgments}. It is a pleasure to thank  Dan Garisto and Phillip Ball 
for posing several questions \cite{Ga22,Ba22}
which inspired this work. We are also grateful to Albert Rico 
for numerous fruitful discussions and to 
Marco Enriquez for his encouragement and patience.

This work was partially funded by Narodowe Centrum Nauki 
under the Maestro grant No.\ DEC-2015/18/A/ST2/00274, 
by Foundation for Polish Science 
under the Team-Net project No.\ POIR.04.04.00-00-17C1/18-00,
by the Center for Quantum Information Theory in Matter and Spacetime, IIT Madras, the Department of
Science and Technology, Govt.\ of India, under Grant No.\
DST/ICPS/QuST/Theme-3/2019/Q69, 
ERC AdG NOQIA, Agencia Estatal de Investigacion (the R\&D project CEX2019-000910-S, funded by MCIN/AEI/10.13039/501100011033, Plan National FIDEUA PID2019-106901GB-I00, FPI), Fundaci\'{o} Privada Cellex, Fundaci\'{o} Mir-Puig, and from Generalitat de Catalunya (AGAUR Grant No.\ 2017 SGR 1341, CERCA program). 

\appendix
\section{Basic notions of the quantum theory}
\label{basics}

The purpose of this Appendix is to explain the nomenclature of quantum information from the main body of the paper.
We shall assume full knowledge about quantum states, i.e.\ that they are pure.
Quantum-mechanical description of the microworld uses vectors of length $N$ (from a Hilbert space $\mathcal{H}_N$ of dimension $N$) to describe quantum states with $N$ levels, e.g.\ 
\begin{equation}\label{eq:computational_basis_qubit}
    \ket{0} =
    \begin{pmatrix}
    1 \\ 
    0
    \end{pmatrix},
    \quad 
    \ket{1} = 
    \begin{pmatrix}
        1 \\ 
        0
    \end{pmatrix}
\end{equation}
represent two perfectly distinguishable states of a two-level system (qubit). 
Classically, these states are analogous to on/off states of bits in an ordinary computer. 

In order to provide a measure for the distinguishability of two quantum states, we use a product of vectors
\begin{equation}\label{eq:def_product}
    \braket{\phi|\psi} = \sum_i a^*_i b_j,
\end{equation}
where complex numbers $a_i$ and $b_i$ refer to coefficients of the quantum states $\ket{\phi}$ and $\ket{\psi}$
\begin{equation}\label{eq:states_phi_psi}
    \ket{\phi} = 
    \begin{pmatrix}
        a_1 \\
        a_2 \\
        \vdots \\
        a_N
    \end{pmatrix},
    \quad 
    \ket{\psi} = 
    \begin{pmatrix}
        b_1 \\
        b_2 \\
        \vdots \\
        b_N
    \end{pmatrix}.
\end{equation}
If two states are orthogonal (their product is 0), then such states are perfectly distinguishable. 
However, if two states overlap (their product is non-zero) then these states cannot be distinguished with probability 1.

Up till now, to describe states we have implicitly used the special computational basis.
Therefore, an alternative description of the state $\ket{\phi}$ from Eq.~(\ref{eq:states_phi_psi}) is given by a combination of the basis states
\begin{equation}
    \ket{\phi} = \sum_i a_i \ket{i}.
\end{equation}

An example of the computational basis is given by states in Eq.~(\ref{eq:computational_basis_qubit}). 
Their orthogonality means that $\braket{0|1} = 0$.
Generally, all the basis states are orthogonal to each other
\begin{equation}
    \braket{i|j} = \delta_{ij} =    
    \begin{cases}
      0 & \text{if $ i\ne j$},\\
      1 & \text{if $i=j$}.
    \end{cases}     
\end{equation}

However, this choice of the basis is not the only one -- any set of $N$ vectors $\{\ket{v_i}\}_{i=1}^N$ orthogonal to each other will form a proper basis. 
For practical purposes, we usually require them to be normalized, $\braket{v_i|v_i} = 1$. 
A particularly useful example of another basis of a qubit system is formed by $\{\ket{+},\ket{-}\}$
\begin{equation}
    \ket{\pm} = \frac{1}{\sqrt{2}} \big(\ket{0} \pm \ket{1}\big).
\end{equation}

These states are orthogonal to each other ($\braket{+|-} = 0$), and such a basis is unitarily equivalent to the computational basis, as exemplified by matrices with columns given by appropriate states
\begin{equation}
    \begin{pmatrix}
        \ket{+} &
        \ket{-}
    \end{pmatrix} = 
    H
    \begin{pmatrix}
        1 & 0 \\
        0 & 1
    \end{pmatrix} = 
    H
    \begin{pmatrix}
        \ket{0} & \ket{1}
    \end{pmatrix},
\end{equation}
where $H = \frac{1}{\sqrt{2}}\begin{pmatrix}
    1 & 1 \\
    1 & -1
\end{pmatrix}$ denotes a Hadamard matrix.

In the case of a $4$-level system, it is sometimes convenient to divide it into 2 parts of 2 levels each. 
Consequently, we name it a two-qubit system; such a system might exhibit non-classical correlations between its subsystems which we call \emph{entanglement}.
The states from the computational basis of a two-qubit system $\{\ket{0},\ket{1},\ket{2},\ket{3}\}$ are not enough to reveal entanglement; however, famous \emph{Bell states} possess the maximal degree of entanglement.
These states $\{\ket{\alpha_i}\}_{i=1}^4$ form a basis of the two-qubit system with $\braket{\alpha_i|\alpha_j} = \delta_{ij}$;
\begin{equation}\label{eq:Bell_basis}
    \begin{split}
        \ket{\alpha_1} &= \big(\ket{00} +\ket{11}\big)/\sqrt{2}, \\ 
        \ket{\alpha_2} &= \big(\ket{01} + \ket{10}\big)/\sqrt{2}, \\
        \ket{\alpha_3} &= \big(\ket{01} -\ket{10}\big)/\sqrt{2}, \text{ and} \\
           \ket{\alpha_4} &= \big(\ket{00} - \ket{11}\big)/\sqrt{2}.
    \end{split}
\end{equation}

An especially useful operation on matrices is the notion of the partial trace of a matrix $M$ of dimension $d_A d_B$.
It can be defined for $M= M_{d_A} \otimes M_{d_B}$ as
\begin{equation}{\rm Tr}_A (M_{d_B}\otimes M_{d_B}) = ({\rm Tr}\otimes \mathbb{I}_{d_B})(M_{d_A}\otimes M_{d_B})={\rm Tr}M_{d_A}\cdot M_{d_B}
\end{equation}
and 
\begin{equation}{\rm Tr}_B (M_{d_A}\otimes M_{d_B}) = (\mathbb{I}_{d_A}\otimes {\rm Tr})(M_{d_A}\otimes M_{d_B})={\rm Tr}M_{d_B}\cdot M_{d_A},
\end{equation}
where $d_A$ and $d_B$ are dimensions of appropriate subsystems represented by the matrix $M_{d_A}$ or $M_{d_B}$.
Since the partial trace is linear, from the above definition one can obtain the partial trace of any matrix $M=\sum_i M_{Ai} \otimes M_{Bi}$.
The partial trace is widely used across quantum information; in particular, the partial trace of a maximally entangled state is maximally mixed.
In other words, although the state is entirely determined, no information about it is accessible from a perspective of a single subsystem.
For instance, the partial trace of any Bell state from Eq.~(\ref{eq:Bell_basis}) amounts to the identity matrix $\text{Tr}_A \ket{\alpha_i}\bra{\alpha_i} = \mathbb{I}_2/2$.

The Bell basis, defined above, can be conveniently represented by a unitary matrix of order four,
\begin{equation}
\label{bell}
    B= \frac{1}{\sqrt{2}}
    \begin{pmatrix}
        1 & 0 & \ 0 & \ 1  \\
         0 & 1 & \ 1 & \ 0  \\
          0 & 1 & -1 & \ 0  \\
          1 & 0 & \  0 & -1 
    \end{pmatrix} ,
\end{equation}
so that each Bell state reads, $|\alpha_i\rangle =B|i\rangle$
with $i=1$, $2$, $3$, $4$.

This is not the only basis composed of maximally entangled states; another example is given by the following vectors      
\begin{equation}
    \begin{split}
        \ket{\beta_1} &= \big(\ket{00} + \ket{01} + \ket{10} + \ket{11}\big)/2, \\ 
        \ket{\beta_2} &= \big(\ket{00} + \ket{01} - \ket{10} + \ket{11}\big)/2, \\
        \ket{\beta_3} &= \big(\ket{00} - \ket{01} + \ket{10} + \ket{11}\big)/2, \text{ and} \\
        \ket{\beta_4} &= \big(-\ket{00} + \ket{01} + \ket{10} + \ket{11}\big)/2.
    \end{split}
\end{equation}

Finally, an especially important maximally entangled basis of a two-qubit system is given by $\ket{\gamma_i}$ obtained from the green-yellow block of size four plotted in Fig.~3 in the original paper~\cite{RBBRLZ22}:
\begin{equation}
    \begin{split}
        \ket{\gamma_1} &= a\omega\ket{00} +a\omega^{19}\ket{01} + b\omega^{14}\ket{10} +b\omega^{16}\ket{11}, \\ 
        \ket{\gamma_2} &= a\omega\ket{00} +a\omega^{3}\ket{01} + b\omega^{10}\ket{10} +b\omega^{4}\ket{11}, \\
        \ket{\gamma_3} &= b\omega^{4}\ket{00} +b\omega^{18}\ket{01} + a\omega^{3}\ket{10} +a\omega^{9}\ket{11}, \text{ and} \\
        \ket{\gamma_4} &= b\omega^{2}\ket{00} +b\omega^{8}\ket{01} + a\omega^{5}\ket{10} +a\omega^{15}\ket{11}.
    \end{split}
\end{equation}
The coefficients read: $a = \big(5+\sqrt{5} \big)^{-1/2}$ and $b = \big((5+\sqrt{5}) /20\big)^{-1/2}$, while phases $\omega^k$ are given by the twentieth root of unity $\omega = e^{i \pi /10}$, with $k \in \{0,1,..., 19\}$.
Interestingly, the number $b/a=(1+\sqrt{5})/2=\varphi$ is the golden ratio, which prompted us to name the state as the \emph{golden} AME$(4, 6)$ state.
To show that these states form a basis, consider a product of two of these vectors, e.g. $\ket{\gamma_2}$ and $\ket{\gamma_3}$ using Eq.~(\ref{eq:def_product}):
\begin{equation}
\begin{split}
    \braket{\gamma_2|\gamma_3} &= a\omega^{-1}b\omega^{4} +a\omega^{-3}b\omega^{18} + b\omega^{-10}a\omega^{3} +b\omega^{-4}a\omega^{9} = ab\big(\omega^{3} +\omega^{15} + \omega^{13} + \omega^{5}\big)=
    \\
    &= ab \big(\omega^{3} - \omega^{5} - \omega^{3} + \omega^{5}\big)= 0.
\end{split}
\end{equation}
The orthogonality conditions are satisfied due to a proper choice of complex phases $\omega$.
Therefore, the states forming blocks of the golden AME state create a maximally entangled basis of the two-qubit system.
Equivalently, we can say that the states $\{\ket{\gamma_i}\}$ are perfectly distinguishable. 

\section{Full form of the golden AME state}
\label{AME_full_formulas}

Explicit expressions for the row
of matrix $U$, corresponding to the chessboard in Fig. \ref{fig:UG_phase}
with
\begin{equation}
a=\frac{1}{2}\sqrt{1-\frac{1}{\sqrt{5}}}
\ < \
b=\frac{1}{2}\sqrt{1+\frac{1}{\sqrt{5}}}
\ < \ 
c=\frac{1}{\sqrt{2}}\qquad\text{and}\qquad \omega=\exp\frac{i \pi}{10}.
\end{equation}

Each non-zero element in matrix $U$ can be encoded in the senary (base-six) numeral system $\{|j\rangle\}_{j=0}^5$, e.g. $|12\rangle=|1\rangle\otimes|2\rangle\leftrightarrow 9^{\rm th}$ index, or $|55\rangle\leftrightarrow 36^{\rm th}$ index. See~\ref{basics} for a primer for the algebra of quantum states.
On the other hand, one can encode matrix $U$ using chess figures and colors due to the following scheme:
\begin{equation}
\kingB = 0, \,\,\,\,\,
\queenB = 1, \,\,\,\,\,
\knightB = 2, \,\,\,\,\,
\bishopB = 3, \,\,\,\,\,
\rookB = 4, \,\,\,\,\,
\pawnB = 5 
\end{equation}
and for colors:
\begin{equation}
{\bf\color{red}red} = 0, \,\,\,\,\,
{\bf\color{cyan}cyan} = 1, \,\,\,\,\,
{\bf\color{green}green} = 2, \,\,\,\,\,
{\bf\color{magenta}magenta} = 3, \,\,\,\,\,
{\bf\color{blue}blue} = 4, \,\,\,\,\,
\def\primarycolor{yellow}%
\def\secondarycolor{black!50}%
\def\shadowHoffset{.5pt}%
\def\shadowVoffset{-.5pt}%
\shadowfy{yellow} = 5.
\end{equation}
Now $|12\rangle=|{\color{green}\queenB}\rangle$ and
\def\primarycolor{yellow}%
\def\secondarycolor{black!50}%
\def\shadowHoffset{.5pt}%
\def\shadowVoffset{-.5pt}%
 $|55\rangle=|\shadowfy{{\pawnB}}\rangle$.

All rows of $U$ are presented below in the standard Dirac notation and chess notation, cf. Fig.~\ref{fig:UG_phase}. 
Note that for example $|\psi_{00}\rangle$ has three components, in accordance with the fact that the first row in Fig.~\ref{fig:matrix_U} contains exactly three non-zero elements. 
Here, every row of a matrix $U$ is treated as a quantum state $\ket{\psi_{ij}}$.
\smallskip

$|\psi_{00}\rangle =c|10\rangle +a\omega^{3}|43\rangle +b|53\rangle=c|{\color{red}\queenB}\rangle +a\omega^{3}|{\color{magenta}\rookB}\rangle +b|{\color{magenta}\pawnB}\rangle$

$|\psi_{01}\rangle =c|00\rangle +b|43\rangle +a\omega^{7}|53\rangle=c|{\color{red}\kingB}\rangle +b|{\color{magenta}\rookB}\rangle +a\omega^{7}|{\color{magenta}\pawnB}\rangle$

$|\psi_{02}\rangle =c\omega^{17}|01\rangle +b|24\rangle +a\omega^{5}|34\rangle=c\omega^{17}|{\color{cyan}\kingB}\rangle +b|{\color{blue}\knightB}\rangle +a\omega^{5}|{\color{blue}\bishopB}\rangle$

$|\psi_{03}\rangle =c\omega^{19}|11\rangle +a\omega^{5}|24\rangle +b|34\rangle=c\omega^{19}|{\color{cyan}\queenB}\rangle +a\omega^{5}|{\color{blue}\knightB}\rangle +b|{\color{blue}\bishopB}\rangle$

$|\psi_{04}\rangle =b\omega^{14}|25\rangle\! + \! a\omega^{15}|35\rangle +a\omega^{18}|42\rangle +b\omega^{3}|52\rangle=b\omega^{14}|\shadowfy{{\knightB}}\rangle +a\omega^{15}|\shadowfy{{\bishopB}}\rangle +a\omega^{18}|{\color{green}\rookB}\rangle +b\omega^{3}|{\color{green}\pawnB}\rangle$

$|\psi_{05}\rangle =a\omega|25\rangle +b\omega^{12}|35\rangle +b\omega^{3}|42\rangle +a\omega^{18}|52\rangle=a\omega|\shadowfy{{\knightB}}\rangle +b\omega^{12}|\shadowfy{{\bishopB}}\rangle +b\omega^{3}|{\color{green}\rookB}\rangle +a\omega^{18}|{\color{green}\pawnB}\rangle$

$|\psi_{10}\rangle =c\omega^{10}|23\rangle +c\omega^{10}|50\rangle=c\omega^{10}|{\color{magenta}\knightB}\rangle +c\omega^{10}|{\color{red}\pawnB}\rangle$

$|\psi_{11}\rangle =c\omega^{6}|33\rangle +c|40\rangle=c\omega^{6}|{\color{magenta}\bishopB}\rangle +c|{\color{red}\rookB}\rangle$

$|\psi_{12}\rangle =a\omega^{2}|04\rangle +b\omega^{5}|14\rangle +c\omega^{7}|41\rangle=a\omega^{2}|{\color{blue}\kingB}\rangle +b\omega^{5}|{\color{blue}\queenB}\rangle +c\omega^{7}|{\color{cyan}\rookB}\rangle$

$|\psi_{13}\rangle =b\omega|04\rangle +a\omega^{14}|14\rangle +c\omega^{19}|51\rangle=b\omega|{\color{blue}\kingB}\rangle +a\omega^{14}|{\color{blue}\queenB}\rangle +c\omega^{19}|{\color{cyan}\pawnB}\rangle$

$|\psi_{14}\rangle =b\omega^{4}|05\rangle +a\omega^{9}|15\rangle +a\omega^{2}|22\rangle +b\omega^{13}|32\rangle=b\omega^{4}|\shadowfy{{\kingB}}\rangle +a\omega^{9}|\shadowfy{{\queenB}}\rangle +a\omega^{2}|{\color{green}\knightB}\rangle +b\omega^{13}|{\color{green}\bishopB}\rangle$

$|\psi_{15}\rangle =a\omega^{3}|05\rangle +b\omega^{18}|15\rangle +b\omega^{19}|22\rangle +a|32\rangle=a\omega^{3}|\shadowfy{{\kingB}}\rangle +b\omega^{18}|\shadowfy{{\queenB}}\rangle +b\omega^{19}|{\color{green}\knightB}\rangle +a|{\color{green}\bishopB}\rangle$

$|\psi_{20}\rangle =c\omega^{2}|31\rangle +c\omega^{13}|44\rangle=c\omega^{2}|{\color{cyan}\bishopB}\rangle +c\omega^{13}|{\color{blue}\rookB}\rangle$

$|\psi_{21}\rangle =c\omega^{2}|21\rangle +c\omega^{7}|54\rangle=c\omega^{2}|{\color{cyan}\knightB}\rangle +c\omega^{7}|{\color{blue}\pawnB}\rangle$

$|\psi_{22}\rangle =c\omega^{19}|12\rangle +a\omega^{12}|45\rangle +b\omega|55\rangle=c\omega^{19}|{\color{green}\queenB}\rangle +a\omega^{12}|\shadowfy{{\rookB}}\rangle +b\omega|\shadowfy{{\pawnB}}\rangle$

$|\psi_{23}\rangle =c\omega^{5}|02\rangle +b\omega^{15}|45\rangle +a\omega^{14}|55\rangle=c\omega^{5}|{\color{green}\kingB}\rangle +b\omega^{15}|\shadowfy{{\rookB}}\rangle +a\omega^{14}|\shadowfy{{\pawnB}}\rangle$

$|\psi_{24}\rangle =a\omega|03\rangle +b\omega^{4}|13\rangle +c\omega^{8}|20\rangle=a\omega|{\color{magenta}\kingB}\rangle +b\omega^{4}|{\color{magenta}\queenB}\rangle +c\omega^{8}|{\color{red}\knightB}\rangle$

$|\psi_{25}\rangle =b\omega^{10}|03\rangle +a\omega^{3}|13\rangle +c\omega^{16}|30\rangle=b\omega^{10}|{\color{magenta}\kingB}\rangle +a\omega^{3}|{\color{magenta}\queenB}\rangle +c\omega^{16}|{\color{red}\bishopB}\rangle$

$|\psi_{30}\rangle =b\omega^{10}|01\rangle +a\omega^{15}|11\rangle +a\omega^{4}|24\rangle +b\omega^{7}|34\rangle=b\omega^{10}|{\color{cyan}\kingB}\rangle +a\omega^{15}|{\color{cyan}\queenB}\rangle +a\omega^{4}|{\color{blue}\knightB}\rangle +b\omega^{7}|{\color{blue}\bishopB}\rangle$

$|\psi_{31}\rangle =a\omega^{5}|01\rangle +b|11\rangle +b\omega^{17}|24\rangle +a\omega^{10}|34\rangle=a\omega^{5}|{\color{cyan}\kingB}\rangle +b|{\color{cyan}\queenB}\rangle +b\omega^{17}|{\color{blue}\knightB}\rangle +a\omega^{10}|{\color{blue}\bishopB}\rangle$

$|\psi_{32}\rangle =a|25\rangle +b\omega^{15}|35\rangle +b\omega^{14}|42\rangle +a\omega^{13}|52\rangle=a|\shadowfy{{\knightB}}\rangle +b\omega^{15}|\shadowfy{{\bishopB}}\rangle +b\omega^{14}|{\color{green}\rookB}\rangle +a\omega^{13}|{\color{green}\pawnB}\rangle$

$|\psi_{33}\rangle =b\omega^{15}|25\rangle +a|35\rangle +a\omega^{7}|42\rangle +b\omega^{16}|52\rangle=b\omega^{15}|\shadowfy{{\knightB}}\rangle +a|\shadowfy{{\bishopB}}\rangle +a\omega^{7}|{\color{green}\rookB}\rangle +b\omega^{16}|{\color{green}\pawnB}\rangle$

$|\psi_{34}\rangle =a\omega|00\rangle +b\omega^{16}|10\rangle +b\omega^{10}|43\rangle +a\omega^{5}|53\rangle=a\omega|{\color{red}\kingB}\rangle +b\omega^{16}|{\color{red}\queenB}\rangle +b\omega^{10}|{\color{magenta}\rookB}\rangle +a\omega^{5}|{\color{magenta}\pawnB}\rangle$

$|\psi_{35}\rangle =b\omega^{14}|00\rangle \! + \! a\omega^{19}|10\rangle +a\omega^{5}|43\rangle +b\omega^{10}|53\rangle=b\omega^{14}|{\color{red}\kingB}\rangle +a\omega^{19}|{\color{red}\queenB}\rangle +a\omega^{5}|{\color{magenta}\rookB}\rangle +b\omega^{10}|{\color{magenta}\pawnB}\rangle$

$|\psi_{40}\rangle =c|05\rangle +b\omega^{7}|22\rangle +a|32\rangle=c|\shadowfy{{\kingB}}\rangle +b\omega^{7}|{\color{green}\knightB}\rangle +a|{\color{green}\bishopB}\rangle$

$|\psi_{41}\rangle =c|15\rangle +a\omega^{10}|22\rangle +b\omega^{13}|32\rangle=c|\shadowfy{{\queenB}}\rangle +a\omega^{10}|{\color{green}\knightB}\rangle +b\omega^{13}|{\color{green}\bishopB}\rangle$

$|\psi_{42}\rangle =c|23\rangle +c\omega^{10}|50\rangle=c|{\color{magenta}\knightB}\rangle +c\omega^{10}|{\color{red}\pawnB}\rangle$

$|\psi_{43}\rangle =c\omega^{10}|33\rangle +c\omega^{14}|40\rangle=c\omega^{10}|{\color{magenta}\bishopB}\rangle +c\omega^{14}|{\color{red}\rookB}\rangle$

$|\psi_{44}\rangle =b\omega^{2}|04\rangle +a\omega^{15}|14\rangle +c\omega^{10}|51\rangle=b\omega^{2}|{\color{blue}\kingB}\rangle +a\omega^{15}|{\color{blue}\queenB}\rangle +c\omega^{10}|{\color{cyan}\pawnB}\rangle$

$|\psi_{45}\rangle =a\omega^{5}|04\rangle +b\omega^{8}|14\rangle +c|41\rangle=a\omega^{5}|{\color{blue}\kingB}\rangle +b\omega^{8}|{\color{blue}\queenB}\rangle +c|{\color{cyan}\rookB}\rangle$

$|\psi_{50}\rangle =a\omega^{10}|02\rangle +b\omega^{5}|12\rangle +b\omega^{9}|45\rangle +a\omega^{16}|55\rangle=a\omega^{10}|{\color{green}\kingB}\rangle +b\omega^{5}|{\color{green}\queenB}\rangle +b\omega^{9}|\shadowfy{{\rookB}}\rangle +a\omega^{16}|\shadowfy{{\pawnB}}\rangle$

$|\psi_{51}\rangle =b\omega^{15}|02\rangle +a|12\rangle +a\omega^{16}|45\rangle +b\omega^{13}|55\rangle=b\omega^{15}|{\color{green}\kingB}\rangle +a|{\color{green}\queenB}\rangle +a\omega^{16}|\shadowfy{{\rookB}}\rangle +b\omega^{13}|\shadowfy{{\pawnB}}\rangle$

$|\psi_{52}\rangle =b\omega^{14}|03\rangle +a\omega^{7}|13\rangle +c\omega^{10}|30\rangle=b\omega^{14}|{\color{magenta}\kingB}\rangle +a\omega^{7}|{\color{magenta}\queenB}\rangle +c\omega^{10}|{\color{red}\bishopB}\rangle$

$|\psi_{53}\rangle =a\omega^{3}|03\rangle +b\omega^{6}|13\rangle +c|20\rangle=a\omega^{3}|{\color{magenta}\kingB}\rangle +b\omega^{6}|{\color{magenta}\queenB}\rangle +c|{\color{red}\knightB}\rangle$

$|\psi_{54}\rangle =c|31\rangle +c\omega|44\rangle=c|{\color{cyan}\bishopB}\rangle +c\omega|{\color{blue}\rookB}\rangle$

$|\psi_{55}\rangle =c\omega^{16}|21\rangle +c\omega^{11}|54\rangle=c\omega^{16}|{\color{cyan}\knightB}\rangle +c\omega^{11}|{\color{blue}\pawnB}\rangle$

\medskip

Simple chess generator and other scripts related to the golden AME$(4,6)$ state can be found on Github~\cite{github}. Note that the matrix
analyzed in the original paper \cite{RBBRLZ22} 
and visualized therein by $36$ cards corresponds to the
partially transposed matrix $U^{\rm \Gamma}$. See also~\ref{app:PRL_BELL}.
It is interesting to observe that an AME state of a heterogenic
system $2 \times 3 \times 3 \times 3$ found in \cite{HESG18}
has a slightly similar structure
with coefficients given by roots of an algebraic equation.

\section{Visual explanation of quantum conditions for OLS}\label{app:visual_ptrace}

The colors in Fig.~\ref{fig:UG_braz} and Fig.~\ref{fig:UG_phase}
are especially attuned.
Note that they form two triplets
of primary colors: {\bf\color{red}red},
{\bf\color{green}green}, {\bf\color{blue}blue} and their appropriate complements:
{\bf\color{cyan}cyan}, {\bf\color{magenta}magenta}, and 
\def\primarycolor{yellow}%
\def\secondarycolor{black!50}%
\def\shadowHoffset{.5pt}%
\def\shadowVoffset{-.5pt}%
\shadowfy{yellow}.
Such a configuration is very helpful in translating the formulas {\bf B'} and {\bf C'}
into a visual explanation, because
we can easily infer -- without cumbersome calculations -- which combinations of the chess figures compensate, exactly as it is in the case of the color domain.

Consider for example formula {\bf B'} and the first two rows in Fig.~\ref{fig:UG_phase}. Since a chess figure of a given color simply encodes the position of the non-zero element in the unitary matrix $U$, (see~\ref{AME_full_formulas}), it is easy to understand that the only contributions to the partial trace over the second subsystem will occur when two figures of the same color appear in both rows. 
Namely, the {\bf\color{red}red} queen, king, pawn, and rook will combine with their {\bf\color{cyan}cyan} counterparts. To confirm this formally one must 
consider full formulas including phases and amplitudes:
\begin{equation}
{c^2w^{-10}|\color{red}\queenB}\rangle\langle{\color{red}\pawnB}|+
c^2|{\color{red}\kingB}\rangle\langle{\color{red}\rookB}|
=-\frac{1}{2}|{\color{red}\queenB}\rangle\langle{\color{red}\pawnB}|
+\frac{1}{2}|{\color{red}\kingB}\rangle\langle{\color{red}\rookB}|
\end{equation}
and
\begin{equation}
c^2|{\color{cyan}\queenB}\rangle\langle{\color{cyan}\pawnB}|+
{c^2w^{10}|\color{cyan}\kingB}\rangle\langle{\color{cyan}\rookB}|
=\frac{1}{2}|{\color{cyan}\queenB}\rangle\langle{\color{cyan}\pawnB}|
-\frac{1}{2}|{\color{cyan}\kingB}\rangle\langle{\color{cyan}\rookB}|.
\end{equation}
After tracing out the colors one obtains
\begin{equation}
-\frac{1}{2}|\queenB\rangle\langle\pawnB|
+\frac{1}{2}|\kingB\rangle\langle\rookB|
+\frac{1}{2}|\queenB\rangle\langle\pawnB|
-\frac{1}{2}|\kingB\rangle\langle\rookB|=0.
\end{equation}

As a more complicated example, let us consider the last two rows in Fig.~\ref{fig:UG_phase} to see how the 
\shadowfy{yellow}
figures coincide with the {\bf\color{blue}blue} ones.
We have
\begin{equation}
    a c w^{-16} |\shadowfy{{\kingB}}\rangle\langle\shadowfy{{\pawnB}}|
    +    a cw^{-16} |\shadowfy{{\queenB}}\rangle\langle\shadowfy{{\rookB}}|
    +   b cw^{-9} |\shadowfy{{\kingB}}\rangle\langle\shadowfy{{\rookB}}|
        + b cw^{-13} |\shadowfy{{\queenB}}\rangle\langle\shadowfy{{\pawnB}}|
        \label{yellowFigs}
\end{equation}
and
\begin{equation}
    a c w^{-6} |{\color{blue}\kingB}\rangle\langle{\color{blue}\pawnB}|
    + a c w^{14} |{\color{blue}\queenB}\rangle\langle{\color{blue}\rookB}|
    + b c w |{\color{blue}\kingB}\rangle\langle{\color{blue}\rookB}|
     + b c w^{-3} |{\color{blue}\queenB}\rangle\langle{\color{blue}\pawnB}|.
     \label{blueFigs}
\end{equation}
Again, after tracing out the color, we see that all terms in~\eqref{yellowFigs} and~\eqref{blueFigs} can be coupled so that the coefficients form antipodal complex numbers on the two circles of radii $ac$ and $bc$, and, when combined, they all vanish. One can repeat this examination for all possible complementary colors in every possible pair of rows.

For the partial trace over the first subsystem in {\bf B'}, the interpretation is slightly different, but one can just exchange the chess figures with colors and the explanation will remain the same.

Finally, exactly the same reasoning is applicable for formulas in {\bf C'}.
However, this time we must consider pairs of figures instead of colors!
Take, for example, the two first rows in Fig.~\ref{fig:UG_phase} and the pair
of $(\bishopB,\knightB)$. We have
\begin{equation}
c^2 |{\color{cyan}\bishopB}\rangle\langle{\color{cyan}\knightB}|
+ b^2 w^{-10} |{\color{blue}\bishopB}\rangle\langle{\color{blue}\knightB}|
+ a^2 w^{-10} |{\color{green}\bishopB}\rangle\langle{\color{green}\knightB}|=...
\end{equation}
Remembering the Pythagorean relation for the amplitudes, we see that
after tracing out the figures, the above formula gives zero
\begin{equation}
   ...= {\color{cyan}c^2} - {\color{blue}\,b^2} - {\color{green}\,a^2}=0.
\end{equation}
As in the previous case, we can check all appropriately coupled figures in all different columns of the array in Fig.~\ref{fig:UG_phase}.
Note that the pair $(\knightB, \bishopB)\neq(\bishopB,\knightB)$, thus
we must consider all six possibilities individually:
$(\kingB,\queenB)$, $(\queenB,\kingB)$, $(\knightB,\bishopB)$,
$(\bishopB,\knightB)$, $(\rookB,\pawnB)$, and $(\pawnB,\rookB)$.

Having gained a little practice, one can immediately predict which elements
in Fig.~\ref{fig:UG_braz} or Fig.~\ref{fig:UG_phase} compensate in partial tracing.
However, this does not directly apply to other variants of the state, especially those presented
in Fig.~\ref{fig:UG_UGR}. That is one of the reasons for which we consider such a particular representation of the AME$(4,6)$ state in the form of matrix~$U$.

\section{Block diagonal form of AME and Bell bases}\label{app:PRL_BELL}

\begin{figure}[ht!]
\center
a) \! \! \includegraphics[width=2.2in]{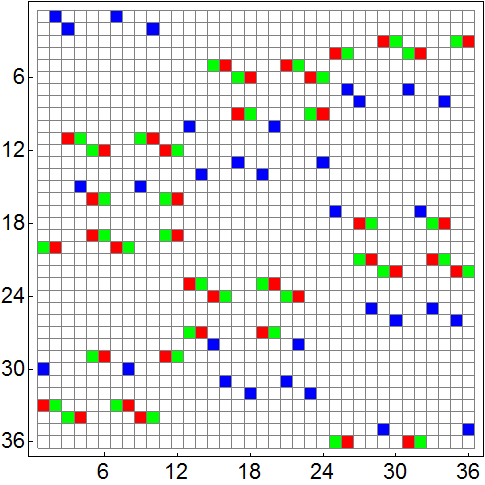} \hskip 0.5cm
b) \!\! \includegraphics[width=2.2in]{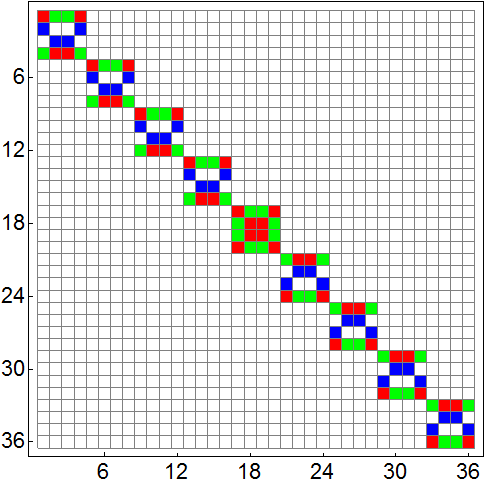}
\caption{a)  Partially transposed matrix $U^{\rm\Gamma}$ corresponding to the chessboard shown 
in  Fig.~\ref{fig:UG_UGR} was already published in ~\cite{RBBRLZ22} 
under the name $\mathcal{U}_{36}$.
b) Its block diagonal form $U^{\rm\Gamma}_{\rm block}$, where each
$4\times 4$ block forms a (maximally entangled) Bell basis for two qubits.
Analogously to Fig.~\ref{fig:matrix_U}, we present only absolute values
where red corresponds to $a$, green to $b$ and blue to $c$ amplitude, respectively.
Note that the block diagonal matrix  $U^{\rm\Gamma}_{\rm block}$ no longer fulfills the properties of
an AME state.
}
\label{fig:matrix_PRL}
\end{figure}

\begin{figure}[ht!]
\center
a) \! \! \includegraphics[width=2.2in]{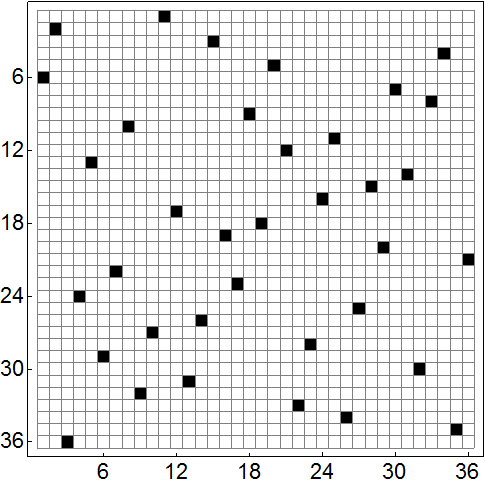} \hskip 0.5cm
b) \!\! \includegraphics[width=2.2in]{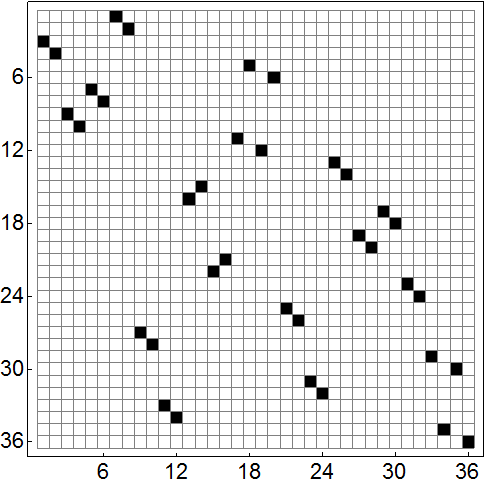}
\caption{a) Left and b) right permutation matrices $P_1$ and $P_2$
defined in (\ref{permut}), which   bring $U^{\rm\Gamma}$
to the block diagonal form presented in Fig.~\ref{fig:matrix_PRL}b.
}
\label{fig:P1P2_for_PRL}
\end{figure}

The matrix $U^{\rm\Gamma}$, 
visualized by the chessboard shown in Fig. \ref{fig:UG_UGR}b,
can be rearranged to a block diagonal form $U^{\rm\Gamma}_{\rm block}$ where nine $4\times 4$
blocks form maximally entangled Bell bases, see Fig.~\ref{fig:matrix_PRL}b.
Two permutation matrices that bring $U^{\rm\Gamma}$ to $P_1U^{\rm\Gamma}P_2=U^{\rm\Gamma}_{\rm block}$, shown in Fig.~\ref{fig:P1P2_for_PRL}, can be written as two vectors consisting of $36$ integers,
\begin{align}
P_1&=[6, 2, 36, 24, 13, 29, 22, 10, 32, 27, 1, 17, 31, 26, 3, 19, 23, 9,\nonumber \\
&18, 5, 12, 33, 28, 16, 11, 34, 25, 15, 20, 7, 14, 30, 8, 4, 35, 21],
\nonumber
\\
\nonumber\\
P_2&=[3, 4, 9, 10, 7, 8, 1, 2, 27, 28, 33, 34, 16, 15, 22, 21, 11, 5, 12, \nonumber \\
&6, 25, 26, 31, 32, 13, 14, 19, 20, 17, 18, 23, 24, 29, 35, 30, 36],
\label{permut}
\end{align}

where non-zero element indicates the position of unity in appropriate column of $P_j$ for $j\in\{1,2\}$.

\end{document}